\documentclass[aps,prc,nofootinbib,twocolumn,10pt,showpacs]{revtex4-1}

\usepackage{graphicx}
\usepackage{epsfig}
\usepackage{amsmath}
\usepackage{xcolor}
\usepackage[normalem]{ulem}  

\newcommand{\midtilde}{\raisebox{-0.25\baselineskip}{\textasciitilde}}

\begin{document}

\title{Non-congruence of the nuclear liquid-gas and 
deconfinement phase transitions}

\author{Matthias Hempel}
\email{matthias.hempel@unibas.ch}
\affiliation{Department of Physics, University of Basel, Basel, Switzerland}

\author{Veronica Dexheimer}
\email{vdexheim@kent.edu}
\affiliation{Department of Physics, Kent State University, Kent OH, USA}

\author{Stefan Schramm}
\email{schramm@th.physik.uni-frankfurt.de}
\affiliation{FIAS, Johann Wolfgang Goethe University, Frankfurt, Germany}
 
\author{Igor Iosilevskiy}
\email{iosilevskiy@gmail.com}
\affiliation{Joint Institute for High Temperature of RAS, Moscow, Russia}
\affiliation{Moscow Institute of Physics \& Technology (State University),
Moscow, Russia}

\date{\today}

\begin{abstract}
First order phase transitions (PTs) with more than one globally 
conserved charge, so-called non-congruent PTs, have characteristic differences
compared to congruent PTs 
(e.g., dimensionality of phase diagrams, 
location and properties of critical points and endpoints).
In the present article we investigate the non-congruence of 
the nuclear liquid-gas PT at sub-saturation densities and the deconfinement PT
at high densities and/or temperatures in Coulomb-less models, relevant for
heavy-ion collisions and neutron stars.
For the first PT, we use the FSUgold relativistic
mean-field model and for the second one the relativistic chiral SU(3) model. The
chiral SU(3) model is one of the few models for the
deconfinement PT, which contains quarks and
hadrons in arbitrary proportions (i.e.\ a ``solution'') and gives a continuous
transition from pure hadronic to pure quark matter above a critical point. The
study shows the universality of the applied concept of non-congruence for the
two PTs with an upper critical point, and illustrates
the different typical scales involved. In addition,
we find a principle 
difference between the liquid-gas and the 
deconfinement PTs: in contrast to the ordinary
Van-der-Waals-like PT, the phase coexistence line of the
deconfinement PT has a negative slope in the
pressure-temperature plane. 
As another qualitative difference we find that the
non-congruent features of the
deconfinement PT become vanishingly small around the critical point.
\end{abstract}

\maketitle
\section{Introduction}
Nuclear matter is expected to undergo two different major phase transitions
(PTs): the liquid-gas phase transition (LGPT) of nuclear matter at
sub-saturation
densities and moderate temperatures and the deconfinement and chiral symmetry
restoration PT at high densities and/or temperatures. For
convenience we will call the latter also the quark hadron phase transition
(QHPT) or QCD PT. These two PTs are actively discussed in the
contexts of heavy-ion collisions and astrophysics. The latter
includes the interior of compact stars, i.e., neutron stars (NS) or 
so-called hybrid stars which have quark matter in their
core.

Various effective models for nuclear matter which are constrained by
properties of nuclei have shown that the LGPT of bulk uniform nucleonic matter,
i.e.\ consisting of neutrons and protons without Coulomb
interactions, is of first order (see Refs.~\cite{fiorilla12,soma09} for two 
recent examples of microscopic models). 
Furthermore, there is also experimental evidence
for this from intermediate-energy heavy-ion collisions
\cite{pochodzalla95,elliott02,bonnet09}. On the other hand,
for smaller systems the LGPT is also found to have critical
behavior \cite{ma97,ma05}. 

For the QHPT the situation is more uncertain.
Ab-initio solutions of QCD exist only for very high densities 
and/or temperatures \cite{Kurkela:2010yk,2010PhRvD..81j5021K,Vuorinen:2003fs,Andersen:2010wu,
Andersen:2011sf}. Simulations on the lattice have shown that the QHPT is a smooth
crossover at vanishing density. Unfortunately their use at finite densities
is problematic because they suffer from the so-called 
``sign problem''. It is a numerical problem found in quantum
mechanical systems of fermions which comes from
the fact that at finite chemical potential the fermion determinant
is complex (see Refs.~\cite{Philipsen:2008zz,fukushima2013} and references 
therein for details). As a consequence, 
effective models for QCD matter have to be used, resulting in
different varieties of possible QCD phase diagrams 
\cite{blaschke05,Baym:2008me,2012APS..APRD10001P,Yamamoto:2007ah,Yamamoto:2008zw,
Abuki:2010jq,Kneur:2010yv,Steinheimer:2009nn,Steinheimer:2011ea,Lourenco:2012dx,
Lourenco:2012yv,McLerran:2007qj,Herbst:2010rf,Hatsuda:2006ps,Bratovic:2012qs,
Sasaki:2010bp,Andronic:2009gj,mintz12}. Many of these models predict that the
QCD PT at low temperatures is of first order like the LGPT, but
some also predict a cross-over transition in this regime. 
In the present investigation we assume 
that both the LGPT and the QHPT are of first order, 
and concentrate on the detailed thermodynamic aspects of the two phase
transitions and especially their non-congruent features. 

A non-congruent phase
transition (NCPT) naturally occurs for a first-order PT with more than 
one globally conserved charge. In this case it becomes possible
that the local concentrations of the charges vary during a phase 
transformation, i.e., the crossing of a phase-coexistence region. This leads to 
qualitative differences compared to congruent PTs. Consider for 
example a phase diagram 
in the temperature--pressure plane. For a given temperature, a NCPT
occurs over a range of pressure, related to a range of local 
concentrations of the charges. For a congruent PT, the equilibrium
conditions can only be fulfilled at a single value of the pressure for each 
temperature. As we will show, also other characteristics 
of PTs depend on the number of globally conserved charges.  
It is the main scope of the present article to identify and discuss 
these non-congruent features.
 
As will be explained below, isospin symmetric matter is an 
``azeotrope'', which means that it leads to a 
congruent PT even though it consists of more than one globally
conserved charge. Consequently, the non-congruent features only become visible
for an isospin asymmetric system, and are thus highly related to the 
isospin asymmetry. 
Phase diagrams of isospin asymmetric matter are of extreme importance for the
complete understanding of QCD and nuclear matter. They are highly related to the
symmetry energy, as explained, e.g., in
Refs.~\cite{DiToro:2009ig,DiToro:2006pq,zhang2012}. Such studies are also used
to analyze the effect of model parameters on the QCD phase diagram
\cite{Shao:2011ij,Liu:2011kc,Shao:2011fk,blaschke13}. The effect of different
isospin/charge assumptions has been studied already extensively in the literature for
the LGPT \cite{BBP,lattimer1978,barranco80,muller95,ducoin06,zhang2012,lavagno12}, 
and also experimentally \cite{sfienti09,mcintosh12}, and for the
QHPT \cite{muller97,Shao:2011ij,Shao:2011fk} as well.
Some authors \cite{muller95,muller97,tatsumi11,aguirre12,lavagno12} have
stated that the LGPT and QHPT changes from first to second order (according
to the Ehrenfest classification) if one goes to an asymmetric system.
This was concluded from the non-standard behavior of thermodynamic quantities
during an isothermal crossing of the two-phase region. One of the main
statements of the present paper is that this non-standard behavior in asymmetric
matter is the typical manifestation of a non-congruent first-order PT. 

In the present article, the LGPT and the QHPT
are studied for the scenarios of heavy-ion collisions of symmetric
and asymmetric nuclei. For the QHPT, we are also investigating the scenario of
the interior of a neutron, respectively hybrid star. For this purpose we use a relativistic
chiral SU(3) effective model. This model predicts that both
the LG and the QCD PT at low temperatures are of first order. Due to
technical reasons, the chiral SU(3) model is applied only for the QHPT. For the
description of the LGPT we apply the FSUgold relativistic mean-field model. Even
using only one selected theoretical model for the QHPT and one for the
LGPT, our main conclusions are to some extent
model-independent, because the applied thermodynamic concepts are rather
universal.

The structure of this article is as follows: in Sec.~\ref{sec_congruence} we
discuss various aspects of (non-)congruence of PTs in detail. In Sec.~\ref{sec_fsugold} we describe the
effective model used for the calculations of the LGPT, the FSUgold equation of state (EOS). In Sec.~\ref{sec_su3} we
continue with the description of the chiral SU(3) model for the QHPT. In
Sec.~\ref{sec_thermo} we specify our thermodynamic model and setup used for the
two PTs and the different physical systems. In
Sec.~\ref{sec_results} we analyze and compare in detail the results for the
different scenarios, with a focus on the structure of the resulting phase diagrams
and the non-congruent features. In Sec.~\ref{sec_conclusions} we summarize
our main findings and draw conclusions. 

\section{Congruence/non-congruence of phase transitions}
\label{sec_congruence}
\subsection{Definition of non-congruence}
The term ``non-congruent'' (or ``incongruent'') phase transition
(NCPT) denotes the situation of phase coexistence of two (or more) macroscopic
phases with different chemical compositions (see the IUPAC definition
\cite{clark94} and Ref.~\cite{iosilevskiy10}). Such systems are also called
``binary'', ``ternary'', etc., in contrast to ``unary'' systems. 
NCPTs are well known since long ago in many
terrestrial applications as a particular type of PTs (regardless of the
term), e.g.\ in low-temperature solution theory (see e.g.\
Ref.~\cite{landau69}), in the theory of simple liquid mixtures of hydrocarbons
(see e.g.\
Ref.~\cite{reid66}), or in the theory of crystal-fluid and crystal-crystal phase
diagrams in chemical compounds. NCPTs are also known in nuclear
physics \cite{barranco80}, in heavy-ion physics \cite{greiner87},
and also in the physics of compact stars \cite{glendenning92} since
quite some time, but the term ``non-congruent'' has been
introduced to these areas of physics only recently (see below).

The variants of terrestrial NCPT which are the closest ones to LGPT and
QHPT discussed in the present article, are PTs
in high-temperature, chemically reacting and partially ionized plasmas---typical
products of extremely heated chemical compounds. NCPTs were studied thoroughly
for high-temperature uranium-oxygen systems during hypothetical ``severe
accidents'' in the framework of nuclear reactor safety problems
\cite{I1,I2,I3,I4,I5,iosilevskiy03}. The universal
nature of this type of PT and its applicability for most astrophysical objects
was claimed and illustrated in Ref.~\cite{iosilevskiy03}, using the examples of 
(hypothetical) plasma PTs in the interiors of Jupiter and Saturn,
brown dwarfs, and extrasolar planets. The identification that most PTs in
neutron stars are non-congruent, in particular for the QHPT in hybrid stars,
was claimed first at several conferences by I.I.\ and 
then published recently in Ref.~\cite{iosilevskiy10}.
Nowadays, the term ``non-congruent'' PT is
already used in the astrophysical literature  \cite{tatsumi11,maruyama12}.
Our theoretical description of the LGPT and QHPT as non-congruent phase transitions
in the present study is based essentially on experience from terrestrial
applications.

It should be noted that in the
above standard terrestrial definition of NCPTs, 
different ionized states of atoms or molecules are not relevant for the possible
non-congruence, but only the number of chemical elements. 
The additional degree of freedom of ionization does not count
in the definition because one deals with phase coexistence
of two electroneutral macroscopic phases (or a mixture of
several electroneutral macroscopic fragments). For macroscopic phases, 
Coulomb interactions automatically lead to local charge neutrality, 
and thereby suppress this degree of freedom. 
Conversely, for all
thermodynamic systems in the present paper, including those corresponding
to matter in neutron stars, Coulomb interactions
are not taken into account explicitly, in spite of the presence of
charged species (protons, quarks, leptons, etc.). 
This is what we call a ``Coulomb-less'' model description. In such a 
Coulomb-less approach positive and negative 
charges (e.g., nuclei and electrons) play the role of different chemical 
elements. 
In nuclear matter the abundance of chemical elements is typically not conserved,
but only some generalized ``charges'' like baryon number, electric charge, and
possibly also isospin or strangeness.
The generalization of the definition of non-congruence to first-order 
PTs in dense nuclear matter, described as Coulomb-less systems 
with more than one
conserved charge, is thus obvious: phase
coexistence of two (or more)
macroscopic phases with different composition of the charges, including
electric charge.

There is a famous example from the context of neutron stars which illustrates
the definition of non-congruence: in beta-equilibrated, cold neutron stars
baryon number
and total net electric charge (which has to be zero) are two conserved
charges. There are two typical choices for the treatment of charge neutrality 
for PTs of macroscopic phases within the Coulomb-less approximation. In the
first case, one assumes local charge neutrality, with zero net charge in both
phases, and thus one
obtains a congruent PT of a unary system. Because here 
the congruence is enforced by the requirement of local charge neutrality we 
call it more specifically to be a ``forced-congruent'' PT as proposed in \cite{iosilevskiy10}.
In astrophysics this scenario is usually
called the ``Maxwell-PT'', which is then used as a synonym for congruent phase
transitions in
general. In the second case one assumes global charge neutrality.
In this case the two coexisting phases will have electric charge concentrations of opposite 
sign. Consequently the system is binary and the PT is non-congruent
\cite{glendenning92}. In astrophysics this is often called the ``Gibbs-PT'', and
again taken as a synonym for non-congruent PTs in general. 
The classification with respect to ``Gibbs'' or ``Maxwell'' of matter in
supernovae or proto-neutron stars with possibly trapped neutrinos was given in
Ref.~\cite{hempel09}. Nuclear matter in heavy-ion collisions also has more than
one conserved charge, namely net baryon number, net
electric charge and also net weak flavor, respectively isospin, due to
the fast timescales involved. Thus Coulomb-less PTs in heavy-ion collisions will
in general also be non-congruent, see also Ref.~\cite{greiner87}.
Ref.~\cite{sissakian06} addresses experimental consequences of the QHPT as a
non-congruent PT. The previous arguments are valid for both PTs,
LGPT and QHPT, just the typical scales involved and the quantitative 
behavior is different. 

\subsection{Coulomb interactions}
\label{sec_coul}
As mentioned before, it should be stressed that for all
thermodynamic systems in the present paper we are using 
a ``Coulomb-less'' model description. 
Also, surface effects are neglected in our work. As
a consequence, the two-phase mixtures at equilibrium (i.e., not
metastable) within the two-phase regions are
always described as coexistence of two \textit{macroscopic} phases. 

The simplification of Coulomb-less is to some extent reasonable for the
theoretical
description of relativistic heavy-ion collisions, where one has a net
electric charge but
Coulomb energies are small compared to the typical collision
energies.
Furthermore, the long-range nature of
Coulomb
forces could be ignored in view of the
small size of the ensemble of heavy-ion
collisions products. 
However, for the same reason it is questionable
whether the thermodynamic limit is fulfilled or not \cite{chomaz05}. On the other hand, for the
description of nuclears clusters appearing in the nuclear liquid-gas PT of low-energy
heavy-ion collsions, Coulomb and surface energies are in fact crucial. 
Nevertheless, the bulk Coulomb-less treatment gives useful insight into the main
characteristics of the PT.

Matter in neutron stars has to be overall charge neutral in order to be 
gravitationally bound. In this case, Coulomb interactions and corresponding 
surface effects can be included in a more detailed mesoscopic description, 
leading to structured mixed phases. Usually these phases with finite-size
substructures are called the ``pasta phases'' 
\cite{ravenhall83,maruyama05,newton09,watanabe11,Avancini:2012bj,Na:2012td} or
``pasta plasma'' \cite{iosilevskiy10}.
The classification with respect to non-congruence of these 
scenarios is somewhat still an open question \cite{iosilevskiy10}. 
In a strict thermodynamic sense, the state of matter in such mesoscopic 
calculations should
not be seen as the two-phase coexistence of a first-order PT, but rather as a
sequence of single phases with non-uniform substructure.

A very low surface tension between the two phases 
(see Refs.~\cite{Pinto:2012aq,mintz12} for possible calculations of 
the surface tension) would 
lead to a highly dispersed
charged and non-soluble
mixture of micro-fragments of one phase into the other, a mixed phase, 
which also could be called a 
charged ``emulsion''.\footnote{Another term (culinary like ``pasta'',
``spaghetti'', etc.) was proposed for this emulsion-like mixture: 
``milk phase'' i.e.\ highly dispersed mixture of
oil micro-drops in water \cite{iosilevskiy10}.}
We remark that very often in the astrophysical literature,
 matter in the two-phase coexistence region of any PT, including those in 
neutral systems, is generally said to be in a ``mixed phase''. We think it is more 
accurate to denote this as a ``two-phase mixture''  and to reserve the term ``mixed phase'' only for the 
state of matter obtained in the mesoscopic description of PTs in Coulomb systems with a low surface
tension, as described above.

Without a detailed mesoscopic treatment, the effect of Coulomb interactions 
in NSs 
can be estimated by
different assumptions for charge neutrality
 \cite{heiselberg93,maruyama08a,Pagliara:2010qm,yasutake12b},
which we will use in the present study. The assumption of local charge 
neutrality, used in the ``Maxwell-PT'' which was already
introduced above, corresponds to the limit of an infinitely high surface 
tension between the two phases. In terrestrial
plasmas, phase equilibrium of locally charge neutral phases with Coulomb 
forces is denoted more accurately as the
Gibbs-Guggenheim conditions for phase equilibrium,
see e.g.\ Ref.~\cite{iosilevskiy10}. 
Conversely, the usage of global charge neutrality (GCN) 
for macroscopic phases in a Coulomb-less approach can be seen as an 
approximation for the case 
of a vanishing surface tension in the mesoscopic description.
In astrophysics, this is typically called the ``Gibbs-PT'' \cite{glendenning92}.

\subsection{Characteristics of non-congruent PT}
\label{subsec_char}
It was shown in
Refs.~\cite{I1,I2,I3,I4,I5,iosilevskiy03,iosilevskiy10,barranco80,greiner87,glendenning92,muller95} 
and many others, and it will also be shown
below, that non-congruency
significantly changes the properties of all PTs, namely: (A) Significant
impact on the phase transformation dynamics, i.e., a strong
dependence of the PT parameters on the rapidity of the transition
\cite{I4}.
(B) The thermodynamics of PTs becomes more complicated. The
essential changes include:
(1) significant change in properties of the
singular points (critical point first of all) and separation of critical point
and endpoints, such as temperature endpoint, pressure endpoint, etc.
(2)
significant change in the scale of two-phase boundaries in extensive
thermodynamic variables (say $P$-$V$, $V$-$T$, $H$-$T$, etc.) and even in
topology of
all two-phase boundaries in the space of all intensive thermodynamic variables,
i.e., pressure, temperature, specific Gibbs free energy etc. Note, that this is
valid for both types of PTs: with and without a critical point (e.g.
gas-liquid-like PT and crystal-fluid-like PT, correspondingly). One of the most
remarkable consequences of the non-congruence in NCPT is the appearance of a
two-dimensional ``banana-like'' region instead of the well-known one-dimensional
$P$-$T$ saturation curve for ordinary (congruent) PTs (see Fig.~1 in
Ref.~\cite{iosilevskiy10}). The same
should be expected in the plane of the widely used pair of variables:
temperature - baryon chemical potential (see below).
(3) Closely connected to this is the
significant change of the behavior in the two-phase region: i.e. isothermal and
isobaric
crossings of the two-phase region do not longer coincide. The isothermal NCPT
starts
and finishes at different pressures, while the isobaric NCPT starts and finishes
at different temperatures. Basically,
the pressure on an isotherm is monotonically increasing with density.

Aspect (3) of NCPTs is well studied in the context of
neutron stars \cite{glendenning92}. Inside a neutron star, the pressure has to
decrease monotonically with the radius. A congruent PT leads
therefore to a spatial separation of the two coexisting phases, with a
discontinuous jump in density and all extensive thermodynamic
variables at the transition
radius inside the neutron star. Conversely, for a NCPT a spatially extended two-phase coexistence region is
present,
with a continuous change of total
density, total energy density, etc, throughout. We remark that for the LGPT there
exist several works which also have discussed the other characteristic features
of NCPTs and only used a partially different terminology,
see, e.g., Refs.~\cite{barranco80, muller95,lavagno12}. For the QHPT, the
possible non-congruence has not been discussed in such detail as is done here. 

Furthermore, even nowadays publications are still appearing which do not treat
the thermodynamic aspects of non-unary phase
transitions, i.e., the non-congruent features, in a proper way. For example
aspect (3) sometimes led to the conclusion that one has a second-order
PT according to the Ehrenfest classification, as e.g., 
in Refs.~\cite{muller95,muller97,tatsumi11,aguirre12,lavagno12}. 
However, the two coexisting phases
have different order parameters like densities, entropies, asymmetries, etc.,
and, most importantly for our purposes, different generalized ``chemical''
compositions. At the interface between the two macroscopic phases there is a discontinuous
jump of the order parameter and thus the PT is still of
first order. Also the first two aspects of (B) from above are sometimes overseen 
or neglected in the
literature which means that the non-congruence is not fully taken into account 
(compare, e.g., with Ref.~\cite{fiorilla12}).

\subsection{Isospin symmetry, azeotrope}
The isospin symmetry of strong forces plays an important role for the possible
non-congruence of the LGPT and the QHPT. Independently of density and
temperature, isospin symmetric nuclear matter always represents the state with
the lowest thermodynamic potential (neglecting Coulomb interactions and assuming
equal masses of protons and neutrons). Thus the isospin chemical
potential is zero for symmetric nuclear matter. As a consequence, isospin does
not appear as a relevant charge for symmetric nuclear matter because this
degree of freedom is not explored, i.e.\ even in a first-order PT
the involved phases remain symmetric. 
Therefore the LG and QH PTs remain congruent if the
system is exactly symmetric
and if no other globally conserved charge than baryon number is involved
(see also Appendix~\ref{app_azeo}). This is called ``azeotropic'' behavior,
denoted for a system with more than
one conserved charge whose charge ratios cannot be changed by distillation for a certain 
azeotropic composition.
The ensemble of such azeotropic points in the parameter space, e.g., for all temperatures, is 
called an azeotropic curve. Note that the isospin asymmetry of
the hot state of matter in a heavy-ion collision experiment is mainly set by the
initial charge to mass ratio $Z/A$ of the colliding nuclei, e.g.,
$Z/A\simeq0.4$ in Au+Au collisions.

\subsection{Unified EOS}
As mentioned in point (1) in Section \ref{subsec_char}, another consequence of non-congruent phase
transitions is the possible emergence of critical points, which are different
from the points of maximum temperature, pressure, or extremal chemical
potential. To obtain such critical points and endpoints at all, it is necessary
that both of the two involved phases are calculated with the same theoretical
model (``unified'' or ``single'' EOS approach). In other words, one has to use
only one underlying many-body Hamilton operator. 
This is in contrast to a ``two-EOS'' approach, where two different EOS models
are applied for the two phases in coexistence. Such a ``two-EOS'' description
can have several short-comings, as it cannot
contain critical points and endpoints 
(see Appendix~\ref{app_cp2eos}, and standard literature, e.g.,
Ref.~\cite{landau69}) and it does not
give a consistent description of meta-stable or unstable matter in the binodal,
respectively spinodal regions, e.g., for a liquid-gas type PT. 
In summary, in
the ``unified'' EOS approach both coexisting phases are presumed as
isostructural (like gas and liquid) with a possible continuous transition from
one phase into another, while in the two-EOS approach this is impossible.

Almost all studies of the LGPT are based on the unified EOS approach. This also
applies for our investigation of the LGPT with the FSUgold relativistic
mean-field model. Unified EOS approaches for the QHPT are usually either built
with only hadrons or only quarks. Thus they do not give the expected degrees of
freedoms for one of the two phases. Alternatively, often the two-EOS
approach is
applied for the QHPT (see, e.g., Ref.~\cite{muller97}) to have
the right
degrees of freedom. 
On the other hand this approach 
cannot contain all
possible non-congruent
features of the singular points, as explained above. The chiral SU(3) model
used in the present study
is one of the few unified-EOS approaches for the QHPT that contains hadronic as
well as quark degrees of freedom. These can appear, in principle, in arbitrary
proportions (solution-like mixture\footnote{Another term was proposed for this
solution-like mixture: ``vodka phase'', i.e., a solution of spirit in water with
arbitrary proportion \cite{iosilevskiy10}.}) with the
interactions leading to the
correct behavior for low, respectively high, densities and temperatures.
See
Refs.~\cite{Steinheimer:2010ib,Steinheimer:2011ea} for another unified-EOS model
that also contains hadronic and quark degrees of freedom. 
As another exception of a unified-EOS approach for the QHPT with the correct
degrees of freedom there is the EOS of Ref.~\cite{randrup10}, where the two-EOS approach
is transformed into a one-EOS version with the use of a special spline-based
interpolation procedure.

\section{FSUgold RMF model}
\label{sec_fsugold}
For the LGPT of nucleonic matter we apply a
relativistic mean-field (RMF) model. In principle, also the Chiral model could
be used for this, as it also contains the LGPT
\cite{Dexheimer:2009hi}. However, due to the different characteristic scales
involved and for numerical reasons, we use a dedicated model for the
LGPT which occurs at sub-saturation densities. We 
choose the FSUgold
RMF parameterization \cite{todd2005}, because of its excellent description of
matter around and below saturation density and because its neutron matter EOS is in
agreement with recent experimental and observational constraints (see e.g.\
Ref.~\cite{lattimer12}). Its Lagrangian is based on the exchange of the
isoscalar--scalar $\sigma$, the isoscalar--vector $\omega$ and the
isovector--vector $\rho$ mesons between nucleons. Particular for FSUgold, also
the coupling between the $\omega$ and the $\rho$ meson is included. This leads
to a better description of nuclear collective modes, the EOS of asymmetric
nuclear matter, and a different density dependence of the symmetry energy
\cite{pieka07}.  The free parameters of the Lagrangian, the meson masses and
their coupling constants, are determined by fits to experimental data,
more specifically to binding energies and charge radii of a selection
of magic nuclei. 

The only baryonic
degrees of freedom in FSUgold are neutrons and protons. For the typical
densities and temperatures of the LGPT, hyperonic or quark degrees of freedom
are not relevant. Because FSUgold considers only the ``elementary'' particles of
the LGPT but not any compound objects, respectively
bound complexes like light
or heavy nuclei, it belongs to the class of so-called physical descriptions
of PTs.
In this description, all effects of bound
complexes are presumed to be taken into account by the interactions (``non-ideality'') of the
``elementary'' particles (see, e.g., Ref.~\cite{ebeling1976}). However, for a
more detailed description of the nuclear EOS like, e.g., used in
simulations of core-collapse supernovae, the formation of nuclei
and nuclear clusters has to be
incorporated explicitly, see e.g.,
Refs.~\cite{shen98,typel09,botvina08,hempel10,raduta10,hempel11,Avancini:2012bj,
buyuk12} . For high
temperatures, light nuclei like the deuteron or alpha particle are most
important, whereas at low temperatures, heavy and also super-heavy nuclei give
the dominant contribution. If all possible compound objects (i.e.\ nuclei) were
included as a chemical mixture, one would obtain a quasi-chemical
representation, as
e.g.\ done in Ref.~\cite{hempel10}. This can cause substantial changes and even
to a quenching of the liquid-gas phase
transition as a first-order PT, see
also Ref.~\cite{ducoin07b}. However, even in this case one
can use the analogy between the characteristic changes of the nuclear composition
and the behavior of the gas and the liquid phases in a pure thermodynamic
treatment, see e.g.\ Refs.~\cite{bugaev01,botvina08}. It is confirmed in many studies,
that the mean-field without clusterization overestimates the 
region of instability, see, e.g., Refs.~\cite{typel09,roepke83}. 
Because Coulomb-interactions and clusterization are more important in the cold
catalyzed matter of neutron stars than in the hot plasma of heavy-ion
collisions, we are discussing the LGPT only in the latter scenario.

\section{Chiral SU(3) model}
\label{sec_su3}
The non-linear realization of the sigma model
\cite{Papazoglou:1998vr,Bonanno:2008tt} is built on the original linear sigma
model \cite{Papazoglou:1997uw,Lenaghan:2000ey}, including the pseudo-scalar
mesons as the angular parameters for the chiral transformation, to be in better
agreement with nuclear physics results. It is an effective quantum relativistic
model that describes hadrons interacting via meson exchange, similar to the
FSUgold RMF interactions. However, the model is constructed in a chirally
invariant manner as the particle masses originate from interactions with the
medium and, therefore, go to zero at high densities/temperatures. 

The Lagrangian density of the model in the mean-field approximation (all
particles contribute to the global mean-field interactions and are in turn
affected by them), constrained further by astrophysical data, can be found in
Refs.~\cite{Dexheimer:2008ax,Negreiros:2010hk,Dexheimer:2011pz}. In this work,
we are going to use an extension of this model called the Chiral SU(3) model,
that also includes quarks \cite{Dexheimer:2009hi}. 
The Lagrangian density in mean-field approximation reads:
\begin{eqnarray}
&\mathcal L = \mathcal L_{\rm Kin}+\mathcal L_{\rm Int}+\mathcal L_{\rm Self}
+\mathcal L_{\rm SB}-U,& \label{lsu3}
\end{eqnarray}
where besides the kinetic energy term $\mathcal L_{\rm Kin}$ the terms
\begin{eqnarray}
&\mathcal L_{\rm Int}=-\sum_i \bar{\psi_i}[\gamma_0(g_{i\omega}\omega+g_{i\phi}\phi+g_{i\rho}\tau_3\rho)+M_i^*]\psi_i, \; \\
&\mathcal L_{\rm SB}= m_\pi^2 f_\pi\sigma+\left(\sqrt{2}m_k^ 2f_k-\frac{1}{\sqrt{2}}m_\pi^ 2 f_\pi\right)\zeta,
\end{eqnarray}
represent the interactions between baryons (and quarks) 
and vector and scalar mesons, 
and an explicit chiral symmetry breaking term, responsible for producing the masses of
the pseudo-scalar mesons.  
$\mathcal L_{\rm Self}$ contains the self interactions of
scalar and vector mesons, where we refer to Refs.~\cite{Dexheimer:2008ax,
Dexheimer:2009hi} for details. 

Up, down, and strange
quarks and the whole baryon octet are always considered in the above sum over 
$i$, in the entire phase diagram. 
However, the degrees of freedom which are actuallly populated 
change from hadrons to quarks and vice-versa through the
introduction of an extra field $\Phi$ in the effective masses of the baryons and
quarks. The scalar field $\Phi$ is named in analogy to the Polyakov loop
\cite{Fukushima:2003fw} since it also works as the order parameter for
deconfinement. The potential for $\Phi$ reads:
\begin{eqnarray}\label{3}
U=(a_0T^4+a_1\mu_B^4+a_2T^2\mu_B^2)\Phi^2\\ \nonumber
+a_3T_0^4\log{(1-6\Phi^2+8\Phi^3-3\Phi^4)}.
\end{eqnarray}
It was modified from its original form in the PNJL model \cite{Ratti:2005jh,
Roessner:2006xn} in order to also be used to study low temperature and high
density environments (besides high temperature and low density environments). It
is a simple form to extend the original potential to be able to reproduce the
physics of the whole phase diagram. Because $U$ now also depends on
the baryon chemical potential $\mu_B$,
it will provide an extra contribution to the total baryon density. It was shown
in Refs.~\cite{Fukushima:2010pp,Lourenco:2012dx,Lourenco:2012yv} that our choice
for the potential $U(\Phi)$ can also be used in the PNJL model, 
successfully
reproducing QCD features. Note that our finite-temperature calculations
include the heat bath of hadronic and quark quasiparticles and their antiparticles
within the grand canonical potential of the system.
Free pions and kaons are included originally in the model, but neglected
here for simplicity. Further comments about their role in the scenarios which we
consider are given below in Sec.~\ref{scen_and_constr}.

With the Lagrangian above, the particle masses are generated by the scalar
mesons whose mean-field values correspond to the isoscalar-scalar
($\sigma$) and
isovector-scalar ($\delta$) light quark-antiquark condensates as well as the
strange
quark-antiquark condensate ($\zeta$). In addition, there is a small explicit
mass term $M_0$ and the term containing $\Phi$:
\begin{equation} 
M_{B}^*=g_{B\sigma}\sigma+g_{B\delta}\tau_3\delta+g_{B\zeta}\zeta+M_{0_B}+
g_{B\Phi} \Phi^2,
\label{1}
\end{equation}
\begin{equation}
M_{q}^*=g_{q\sigma}\sigma+g_{q\delta}\tau_3\delta+g_{q\zeta}\zeta+M_{0_q}+
g_{q\Phi}(1-\Phi).
\label{2}
\end{equation}
We remark that for FSUgold only the term with the sigma field (with a minus
sign) and a large explicit mass term $M_{0_B}$ equal to the nucleon vacuum
mass, would be present in Eq.~(\ref{1}). For FSUgold, the contribution of the
sigma field is zero in the vacuum and decreases the effective mass for finite
density. In the Chiral SU(3) model, the explicit mass term is much smaller, and
the nucleon mass in the vacuum is generated mainly by the $\sigma$ field
(non-strange chiral condensate). With the increase of density/temperature, the
$\sigma$ field (non-strange chiral condensate) decreases from its high value at
zero density, causing the effective masses of the particles to decrease towards
chiral symmetry restoration. 

The coupling constants of Eqs.~(\ref{lsu3})-(\ref{2}) can be found in
Refs.~\cite{Dexheimer:2008ax,Dexheimer:2009hi}. They were chosen to reproduce 
the vacuum masses of baryons and mesons, nuclear saturation properties,
symmetry energy, hyperon optical potentials,
lattice data as well
as information about the QCD phase diagram from
Refs.~\cite{Ratti:2005jh,
Roessner:2006xn,aoki06,Fodor:2004nz}. The model reproduces a realistic QCD 
phase diagram where at the
critical endpoint a first-order PT line begins. The line is
calibrated to terminate on the zero temperature axis at four times saturation
density for charge-neutral beta-equilibrated matter. In this way we can reproduce a hybrid star containing a quark core. The behavior of
the order parameters and the resulting phase diagrams will be discussed in Sec.~\ref{sec_results}.

The most important aspect of the chiral SU(3) model is
that hadrons are included as quasi-particle degrees of
freedom in a chemical equilibrium mixture with quarks. Therefore, the model
gives a quasi-chemical
representation of the deconfinement PT (so-called ``chemical
picture'' in terms of electromagnetic non-ideal
plasmas, see, e.g., Ref.~\cite{iosilevskiy2000}). As
explained in Sec.~\ref{sec_congruence}, it is very important for our study that this
model contains the right degrees of freedom of low and high densities (namely
hadrons and, respectively, quarks) in arbitrary proportions and 
gives at the same time the deconfinement PT in
a ``unified EOS'' or ``single EOS'' description. 

The assumed full miscibility of hadrons and
quarks is, e.g., in contrast to the underlying picture of simple
quark-bag models. At sufficiently high temperature, this will
also lead to the appearance of quarks soluted in the ``hadronic sea'', i.e., inside what we call the hadronic, respectively confined phase. On the
other hand it is also possible that some hadrons survive being soluted in the ``quark sea'', i.e., in the quark or deconfined phase. Nevertheless,
quarks will always give the dominant contribution in the quark phase, and hadrons in the hadronic phase. This is achieved via the field $\Phi$, which assumes non-zero values with the
increase of temperature/density and, due to its presence in the baryons'
effective masses, suppresses their appearance. On the other hand, the presence of
the $\Phi$ field in the effective mass of the quarks, included with a negative
sign, ensures that they will not be present at low temperatures/densities. The hadronic and the quark
phase are characterized and distinguished from each other by their order
parameters, whereas $\Phi$ is one of them, but also the baryon number density or the asymmetry, as we will show later. 
The identification of the two phases via order parameters can always be done in an unambiguous way whenever one has phase
coexistence. We assume that the inter-penetration of quarks and hadrons in the two phases is physical, and
it is required to obtain the cross-over transition at
low baryon chemical potential.

\section{Thermodynamic setup}
\label{sec_thermo}
\subsection{Definitions} 
\begin{table}
\begin{center}
\caption{\label{tab:qn_single}Definitions of total net quantum numbers or
total net charges which are possibly conserved, depending on the scenario
considered, and the corresponding chemical potentials. $F=F(T,V,B,S,Q)$ denotes
the total free energy which is a function of temperature $T$, volume $V$ and the
total net charges $B, S,$ and $Q$.}
\begin{tabular}{ccc}
\hline
\hline
quantity & definition & chem.\ potential\\
 \hline
baryon number & $B= \sum_i N_i b_i$ & $\mu_B=\left.\frac{\partial F}{\partial
B}\right|_{T,V,S,Q}$ \\
strangeness & $S= \sum_i N_i s_i$ & $\mu_S=\left.\frac{\partial F}{\partial
S}\right|_{T,V,B,Q}$  \\
electric charge & $Q= \sum_i N_i q_i$ & $\mu_Q=\left.\frac{\partial F}{\partial
Q}\right|_{T,V,B,S}$ 
\\
\hline
electric charge fraction & $Y_{Q}=Q/B$ & not used \\
\hline
\hline
\end{tabular}
\end{center}
\end{table}
For the (Coulomb-less) scenarios
we are interested
in, the following three quantum numbers of each particle species $i$ are
relevant: baryon number $b_i$, electric charge number $q_i$ and strangeness
$s_i$. The corresponding values can be found in standard
textbooks, or e.g.~in Ref.~\cite{nakamura10}. The quantum numbers of each
particle species $i$ also set the total net quantum numbers or total
net charges of the thermodynamic system if the total net numbers of particles
$N_i$
of each species $i$ are known. The total net number $N_i$ is the difference
between the 
number of particles and the number of corresponding anti-particles of the whole system.
The possibly conserved total net quantum numbers (which are extensive)
are listed in Table \ref{tab:qn_single}. Very often instead of the electric
charge number $Q$, the intensive charge-to-baryon ratio is used, which is defined in the
last row of the table. For each of the extensive
quantum numbers a corresponding chemical potential
can be
defined. These are listed in the third column of Table
\ref{tab:qn_single}.
Later we will also use the following chemical potential $\tilde \mu$, 
\begin{eqnarray}
\tilde \mu &=& \left.\frac{\partial F}{\partial B}\right|_{T,V,S,Y_Q}
\label{mutilde1}\\
& = &  \mu_B + Y_Q \mu_Q \; , \label{mutilde2}
\end{eqnarray}
which is
equal to the Gibbs free energy per baryon (see
Appendix~\ref{app_gibbsfe}).

\begin{table}
\begin{center}
\caption{\label{tab:consquant}Definitions of net quantum numbers or net charges and
corresponding chemical potentials of the individual phases inside the phase
coexistence region. The free energy $F^I$ of phase $I$ is understood as
$F^I=F^I(T^I,V^I,B^I,S^I,Q^I)$ (analogous definitions for phase
$II$).}
\begin{tabular}{ccc}
\hline
\hline
quantity & definition &  chem.\ pot. \\
 \hline
baryon number & $B^I= \sum_i N_i^I b_i$ 
& $\mu_B^I=\left.\frac{\partial F^I}{\partial B^I}\right|_{T,V^I,S^I,Q^I}$ \\
strangeness & $S^I= \sum_i N_i^I s_i$ 
& $\mu_S^I=\left.\frac{\partial F^I}{\partial S^I}\right|_{T,V^I,B^I,Q^I}$ \\
electric charge & $Q^I= \sum_i N_i^I q_i$ 
& $\mu_Q^I=\left.\frac{\partial F^I}{\partial Q^I}\right|_{T,V^I,B^I,S^I}$ \\
\hline
electric charge fraction & $Y_{Q}^I=Q^I/B^I$ & not used \\
\hline
\hline
\end{tabular}
\end{center}
\end{table}
For a state which is inside the two-phase coexistence region, two spatially
separated macroscopic phases are present. Each phase has its own set of
extensive thermodynamic variables and chemical potentials,
listed in Table \ref{tab:consquant}. The total extensive quantities $F, V, B, S,
Q, N_i$ are given as the linear sums of corresponding quantities of the
coexisting phases. Particle numbers are connected to particle number densities
through the volumes of each phase:
\begin{eqnarray}
& \rho_i^I=N_i^I/V^I \, ,&  \nonumber \\
&\rho_i^{II}=N_i^{II}/V^{II} \, . & 
\end{eqnarray}

\subsection{Scenarios and constraints}
\label{scen_and_constr}
{\setlength{\tabcolsep}{6pt}
\begin{table*}
\begin{center}
\caption{\label{tab:cases}Constraints and particles used in the six different
scenarios. For each particle also the corresponding anti-particle is included. 
If a quantity of Table \ref{tab:consquant} is not listed, its value
is not constrained additionally. Note that the combined conservation of
$B$, $S$ and $Y_Q$ is equivalent to the conservation of baryon number,
strangeness and isospin.}
\begin{tabular}{ccccccc}
\hline
\hline
case & \multicolumn{3}{c}{constraints}& considered particles \\
\hline
 LGS  & $B={\rm const.}$ & $S^I=S^{II}=0$ & $Y_Q=0.5$ & neutrons, protons\\
LGAS  & $B={\rm const.}$ & $S^I=S^{II}=0$ & $Y_Q=0.3$ & neutrons, protons\\
LGAS\_fc & $B={\rm const.}$ & $S^I=S^{II}=0$ & $Y_Q^I=Y_Q^{II}=0.3$ & neutrons,
protons\\
 HIS  & $B={\rm const.}$ & $S^I=S^{II}=0$ & $Y_Q=0.5$ & baryon octet, quarks\\
HIAS  & $B={\rm const.}$ & $S^I=S^{II}=0$ & $Y_Q=0.3$ & baryon octet, quarks\\
HIAS\_fc& $B={\rm const.}$ & $S^I=S^{II}=0$ & $Y_Q^I=Y_Q^{II}=0.3$ &  baryon octet,
quarks\\
NSLCN & $B={\rm const.}$ & - & $Y_Q^I=Y_Q^{II}=0$ & baryon octet, quarks, leptons\\
NSGCN & $B={\rm const.}$ & - & $Y_Q=0$ & baryon octet, quarks, leptons\\
\hline
\hline
\end{tabular}
\end{center}
\end{table*}}
Next we are going to define the different cases of the two PTs
studied in different physical scenarios. An overview of these scenarios is
given in Table
\ref{tab:cases}. We consider PTs in three different physical
systems: the liquid-gas phase transition of nuclear matter (LG), e.g.\ in low
energy heavy-ion collisions, the deconfinement phase transition in high energy 
heavy-ion collisions (HI), and the
deconfinement phase transition in
neutron stars (NS). For the first two scenarios LG and HI we investigate
symmetric (S) nuclear matter with $Y_Q=0.5$, and asymmetric (AS) nuclear matter
with $Y_Q=0.3$. The two different electric charge fractions correspond to
heavy-ion reactions of nuclei with different charge to mass ratios $Z/A$.
For $^{197}$Au, which is commonly used in heavy-ion experiments, one has
$Z/A \simeq 0.4$. However, for peripheral collisions $Y_Q \sim 0.35$ can be
reached at certain stages of the evolution as discussed in
Ref.~\cite{sissakian06}. For all of the asymmetric configurations we also
include a forced-congruent (fc) variant of phase equilibrium
\cite{I1,hempel09,iosilevskiy10}, where the
composition of all conserved charges is forced to be equal in the coexisting
phases in frames of Maxwell conditions. In particular, the charge fraction is
constrained locally.  For the (Coulomb-less) scenarios of NSs, we
investigate the effect of local
(NSLCN) and global charge neutrality (NSGCN). Next we explain the physical
meaning of all of the constraints in more detail. 

We remark again that we consider only coexistence of macroscopic phases
and that we do not consider any Coulomb interactions despite the
significant participation of electrically charged particles, as discussed in
Sec.~\ref{sec_congruence}.
Nevertheless, the electric charge is an important quantity for our
investigations because it is one of the conserved charges which determine
the possible non-congruence. Furthermore, the electric charge is also related
to isospin. Let us assume that also the total net baryon number $B$ and the total
net strangeness $S$ are kept constant, just like in all scenarios of LG and HI. The quantum
numbers of the baryons are directly given by the sum of the quantum numbers of
their constituent quarks. Therefore the total numbers of u-, d- and s-quarks
(free or bound in baryons) are fixed by the total net baryon number $B$,
strangeness $S$ and electric charge number $Q$. If the latter three quantities
are kept constant, the total quark content does not change, i.e.\ flavor is
conserved. This means no weak reactions occur and also the total isospin of the
system is conserved. 

In heavy-ion reactions, the typical timescales are on the order of $10^{-23}$~s
which is much less than weak reaction timescales. Therefore we do not allow for
weak reactions in the cases LG and HI. This is implemented via a fixed value of
$Y_Q$, conservation of baryon number $B$ and conserved total net
strangeness $S=0$. In addition to global conservation of the electric charge in
LGS, LGAS, HIS, and HIAS, we also consider locally constrained charge fractions
in the forced-congruent cases of LGAS\_fc and HIAS\_fc. $S$ is set to zero,
because initially there is no strangeness in the two colliding nuclei. In
principle, there is still the possibility, that one has net strangeness in the
two phases with $S^I=-S^{II}$ which is known as strangeness distillation
\cite{greiner87}. Here we suppress this degree of freedom to avoid a
ternary PT\footnote{With
``ternary'' we mean that one had three globally
conserved charges with three chemical equilibrium conditions.} and set
$S^I=S^{II}=0$ for simplicity. For HIS, HIAS and HIAS\_fc this means that the
total number of strange quarks (free or bound in baryons) is equal to the number
of anti-strange quarks locally and that there is a non-zero strange chemical
potential, with two different values in the two phases. For LGS, LGAS, and
LGAS\_fc strangeness is not relevant at all, because no strange particles 
are considered, but only neutrons and protons. This is appropriate for the
typical low energies where the nuclear LGPT is relevant.
We do not consider leptons in the cases of LG and HI, because they are not
present in the initial configuration and their plasma in the later evolution with
equal amounts of particles and antiparticles would not affect the equilibrium
conditions between baryons and quarks. 

At high temperature, the inclusion of light
real mesons, like pions and kaons, is important for some of the 
thermodynamic quantities (e.g., pressure), since light particles dominate 
in such regime. However, if their interactions with the baryons are negligible,
we do not expect a major influence on the topology of baryonic phase diagrams 
(e.g., in the temperature--baryon chemical potential--plane). In some of the 
scenarios considered, the meson contribution from the two coexisting phases
would cancel exactly or be at least very similar. In this case 
the inclusion of free mesons would correspond only to 
a redefinition or shift of some of the thermodynamic quantities.
Here we are concentrating on the baryonic component and a more detailed 
treatment of mesons is postponed to future work.

In cold neutron stars one typically assumes that all possible reactions have
reached full equilibrium. Weak reactions do not conserve strangeness and,
therefore, it is not listed as a conserved quantity in Table \ref{tab:cases} for
the two cases of neutron stars, NSLCN and NSGCN. In principle, weak
reactions conserve lepton numbers but in cold neutron stars neutrinos can
escape freely and, therefore, the interior lepton numbers are also not conserved.

Finally, electrically
charged matter cannot exist in neutron stars on a macroscopic scale, because
otherwise they would explode, as Coulomb interactions are many orders of
magnitude
stronger than gravity. Thus we also include the lepton
contribution
in form of electrons and muons, which is done easily as they
are well described as ideal Fermi-Dirac gases. We implement electric
charge neutrality in two
different ways, as discussed in the introduction. This is done either via enforced local
charge neutrality (NSLCN), where both macroscopic phases are charge neutral
and Coulomb
forces are absent, or via global charge neutrality (NSGCN) in a Coulomb-less
description, where each of the two phases carries a net electric charge which sum
up to zero.

We remark that the scenarios LG and HI described above could also be taken
as simplified examples of supernova matter, for which one has similar values of
$Y_Q$. On the other hand, supernova matter has to be charge neutral, like matter
in neutron stars, and therefore negatively charged leptons have to be included.
For GCN and the Coulombless approximation, charged leptons would not
influence the behavior of the PTs in cases HI and LG. However, for a
realistic description of the LGPT in supernovae the
Coulombless bulk treatment is not sufficient, and the formation of nuclei and
nuclear clusters has to be taken into account, as noted before.

\subsection{Phase and chemical equilibrium conditions}
\label{sec_eq}
Based on the previous constraints, the equilibrium conditions can be derived.
First we consider the system outside of the phase coexistence region. If
there
are more particle species than conserved charges, conditions for chemical
equilibrium are necessary. The chemical potential $\mu_i$ of particle $i$ is
related to the chemical potentials of the total charges:
\begin{eqnarray}
 \mu_i = b_i \mu_B + s_i \mu_S + q_i \mu_Q  \; , \label{mus}
\end{eqnarray}
which allows to calculate the abundances of all particles, if the values of the
total charges are known. Note that $\mu_S$ is the chemical potential for
strangeness as defined in Table \ref{tab:qn_single}, 
which is different from the chemical potential of the strange quark.
For NSLCN and NSGCN, the
non-conservation of strangeness leads to $\mu_S=0$, which is nothing but the
minimization of the thermodynamic potential with respect to strangeness.

We remark that it is also possible to formulate the equilibrium conditions of
Eq.~(\ref{mus}) by using chemical potentials of three selected particles instead
of the chemical potentials $\mu_B$, $\mu_Q$, and $\mu_S$. We want to give an
example for
this alternative formulation. Taking the
chemical potentials of neutrons, protons and
lambdas as the basic units one obtains from Eq.~(\ref{mus}):
\begin{equation}
\mu_i={b_i} \mu_n+ s_i(\mu_n-\mu_\Lambda) + q_i (\mu_p - \mu_n) \; . 
\label{eq_muyc}
\end{equation}
This sets the chemical potentials of all particles, if $\mu_n$, $\mu_p$, and
$\mu_\Lambda$ are determined according to the external constraints (see
Table~\ref{tab:cases}). For example this would lead to:
\begin{eqnarray}
&\mu_u = \frac13(2 \mu_p -\mu_n)\; ,&\nonumber \\
&\mu_{\Xi^-}= 2 \mu_\Lambda - \mu_p\; .&
\end{eqnarray}
For NSs, where leptons are considered, one would also get:
\begin{eqnarray}
&\mu_e = \mu_{\mu}=\mu_n -\mu_p\; ,
\end{eqnarray}
because of the assumption of non-conservation of the lepton numbers.

Inside the phase coexistence region one has to consider equilibrium conditions
between the two phases. Thermal and mechanical equilibrium are given by:
\begin{eqnarray}
&P= P^I=P^{II}\; ,& \label{condeqg1} \\
&T=T^I=T^{II} \; .& \label{condeqg2}
\end{eqnarray}
Inside each phase, one still has relations analogous to Eq.~(\ref{mus}). They give
the chemical potential of particle $i$ in phase $I$, respectively $II$,
expressed by the local chemical potentials of the charges:
\begin{eqnarray}
&\mu_i^{I} = b_i \mu_B^{I} + s_i \mu_S^{I} + q_i \mu_Q^{I}\; ,& \nonumber \\
&\mu_i^{II} = b_i \mu_B^{II} + s_i \mu_S^{II} + q_i \mu_Q^{II}\; .&
\label{mui_coex}  
\end{eqnarray}

Next, one has the chemical equilibrium conditions between the two
phases. In Coulomb-less sytems, which are equivalent to terrestrial
chemically reacting systems (e.g., Ref.~\cite{iosilevskiy10}), the local
chemical
potentials of all species in coexisting phases must be equal, i.e.,
$\mu_i^I=\mu_i^{II}$, if
no local constraints are applied, according to the
traditional laws of chemical thermodynamics.
In this case $\mu_i^I=\mu_i^{II}$ would also
follow from $\mu_B^I=\mu_B^{II}$, $\mu_Q^I=\mu_Q^{II}$, and
$\mu_S^I=\mu_S^{II}$, and Eqs.~(\ref{mui_coex}).
However, due to the local constraints applied (see Table~\ref{tab:cases}),
the inter-phase chemical equilibrium conditions
depend on the scenario considered, and have to be derived, e.g., by means
of Lagrange-multipliers (see also Ref.~\cite{hempel09}). In the following, we
list the inter-phase chemical equilibrium conditions for the different cases.

\paragraph*{LGS, LGAS, HIS, and HIAS}
\begin{eqnarray}
&\mu_B^I = \mu_B^{II}\; ,&\label{condeq1}\\
&\mu_Q^I =   \mu_Q^{II}\; .& \label{condeq2}
\end{eqnarray}
Note that $\mu_S^I\neq \mu_S^{II}$ in order to have
$S^I=S^{II}=0$.
In the alternative formulation from above, Eqs.~(\ref{condeq1}) and
(\ref{condeq2}) would be equivalent to:
\begin{eqnarray}
&\mu_n^I = \mu_n^{II}\; ,&\\
&\mu_p^I =   \mu_p^{II}\; .&
\end{eqnarray}

\paragraph*{NSLCN}
\begin{eqnarray}
&\mu_B^I = \mu_B^{II}\; ,& \label{mubnslcn}\\
&\mu_S^I =  \mu_S^{II} = 0\; .&
\end{eqnarray}
The latter relation comes from the non-conservation of strangeness and
implies that there is a net strangeness in both of the two
phases. Note that: 
\begin{eqnarray}
\mu_Q^I \neq   \mu_Q^{II} \label{eq_muqneq}\; .
\end{eqnarray}
This means for example:
\begin{eqnarray}
&\mu_p^I \neq   \mu_p^{II} \; ,&\\
&\mu_e^I \neq   \mu_e^{II} \; .&
\end{eqnarray}
We remark that according to the Gibbs-Guggenheim conditions (see for example
Ref.~\cite{iosilevskiy10}), for a macroscopic equilibrium Coulomb system
one should
introduce the \textit{electro-chemical} potential
\cite{guggenheim1933} $\mu_Q^{*I}=\mu_Q^I + V_{\rm Galvani} = \mu_Q^{II}= 
\mu_Q^{*II}$ (relative to an arbitrary constant in uniform Coulomb systems).
With this description, the generalized
electro-chemical potentials of all charged particles would be
equal in the two coexisting macroscopic phases, but this is not used here.

\paragraph*{NSGCN}
\begin{eqnarray}
&\mu_B^I = \mu_B^{II}\; ,&\label{mubnsgcn} \\
&\mu_S^I =  \mu_S^{II}  =0\; ,& \label{musnsgcn} \\ 
&\mu_Q^I =   \mu_Q^{II}\; .& \label{muqnsgcn} 
\end{eqnarray}
So here we have: 
\begin{eqnarray}
&\mu_p^I =   \mu_p^{II} \; ,&\\
&\mu_e^I =   \mu_e^{II} \; .&
\end{eqnarray}

\paragraph*{LGAS\_fc and HIAS\_fc}
\label{sec_lgas_fc} 
Next, we give the
equilibrium
conditions if the local charge fractions are constrained to have the same value,
$Y_Q^I =Y_Q^{II}$ ($=Y_Q$). Because in the considered cases only baryon number
remains as a globally conserved charge, the Maxwell construction for a congruent
PT can be used. It is well known, that for the ``Maxwell'' phase
transition in a neutron star with local charge neutrality and
beta-equilibrium the baryon chemical
potential, which in this case is equivalent to the neutron chemical potential,
has to be equal in the two phases, see Eq.~(\ref{mubnslcn}). For HIAS one
obtains instead the following
inter-phase chemical equilibrium
condition \cite{hempel09}:
\begin{eqnarray}
\tilde \mu^I &=& \tilde \mu^{II}\label{eq_eq1}  \\
\Leftrightarrow  \mu_B^{I} + Y_Q \mu_Q^{I} &=&  \mu_B^{II} + Y_Q \mu_Q^{II}
\label{eq_eq} \; ,
\end{eqnarray}
with the local Gibbs free energy per baryon
\begin{eqnarray}
\tilde \mu^I &=& \left.\frac{\partial F^I}{\partial
B^I}\right|_{T,V^I,S^I,Y_Q^I} \\
& = &  \mu_B^I + Y_Q^I \mu_Q^I \; , \label{def:mut}
\end{eqnarray}
and the analogous expression for $\tilde \mu^{II}$ of phase $II$.
Eq.~(\ref{eq_eq1}) expresses the equality of the specific
Gibbs free energy of the two phases, respectively the Gibbs free energy per
baryon used here (see Appendix \ref{app_gibbsfe}). This is merely the
standard Maxwell construction
for a congruent PT, which is also applicable for the
forced-congruent case.

In general, the
baryon and charge chemical potentials will not be the same for the two phases in
the phase coexistence region, because Eq.~(\ref{eq_eq}) is the only chemical
equilibrium condition for cases LGAS\_fc and HIAS\_fc. For a better comparison
with the non-congruent variants LGAS and HIAS, we will
show the phase diagrams of LGAS, LGAS\_fc, HIAS, and HIAS\_fc not only as a
function of $\mu_B$, but also as a function of $\tilde \mu$. 

\begin{figure}[t]
\begin{center}
\includegraphics[width=\columnwidth]{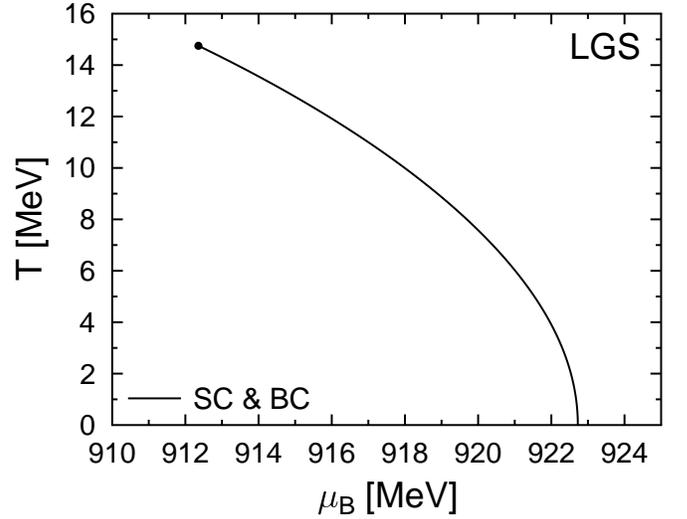}
\caption{\label{fig:tmub_lg_s}Phase diagram in the temperature - baryon chemical
potential plane for case LGS ($Y_Q = 0.5$). The saturation curve
(SC) coincides with the boiling curve (BC). The black dot marks the ordinary
VdW-like critical point (CP).}
\end{center}
\end{figure}

\paragraph*{Total chemical potentials inside the two-phase mixture}
The equilibrium
conditions given above allow one to determine the phase boundaries
and fully specify the properties of the two phases in equilibrium. However,
the non-equality of local chemical potentials due to local constraints
leads to the following complication: in this case it
is not obvious how the total chemical potentials
of the charges in the two-phase mixture (defined analogously to the ones
in Table~\ref{tab:qn_single}, with the local constraints of
Table~\ref{tab:cases} in addition) are related to the 
local chemical potentials of Table~\ref{tab:consquant}, which can have
different values in the two phases. These
relations are derived in Appendix \ref{app_mutot}. We are
not aware that these expressions have been published in the literature before. 

\begin{figure}[t]
\begin{center}
\includegraphics[width=\columnwidth]{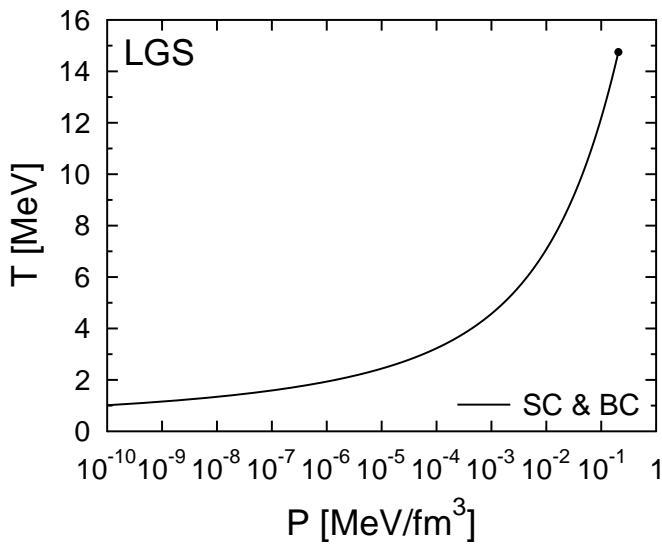}
\caption{\label{fig:pt_lg_s}Phase diagram in the temperature - pressure plane
for case LGS, otherwise the same notation and depiction as in
Fig.~\ref{fig:tmub_lg_s} is used.}
\end{center}
\end{figure}

\section{Results of calculations}
\label{sec_results}
In this section we are showing the results for the phase diagrams of each case
studied, whereas we begin with the LGPT and continue with the QHPT.

\subsection{Nuclear liquid-gas phase transition}
Fig.~\ref{fig:tmub_lg_s} shows the phase diagram of case LGS, i.e.\ for the
liquid-gas phase transition of symmetric nucleonic matter. In principle,
symmetric nuclear matter is a two-component, binary
system of protons and neutrons, respectively baryon number and isospin.
However, the nuclear interactions and isospin symmetry lead to
azeotropic
behavior, i.e., the ratio of protons to
neutrons does not change during
phase coexistence and the two coexisting phases remain
symmetric. The electric charge
chemical potential $\mu_Q$ of symmetric nuclear matter is zero, independently of
density and temperature. Therefore no isospin distillation occurs, i.e.\ there
is no transfer of isospin per baryon, respectively $Y_Q$, between the two
phases. Since $\mu_Q \equiv 0$, the relation of chemical equilibrium with
respect to changes of $Y_Q$, Eq.~(\ref{condeq2}), is automatically fulfilled,
and only Eq.~(\ref{condeq1}) carries relevant information. Consequently,
symmetric nuclear matter behaves like a unary system
and the PT is of congruent type with a
phase-coexistence line in the $T$-$\mu_B$-plane shown in
Fig.~\ref{fig:tmub_lg_s}. This line can be obtained with a Maxwell construction
by the corresponding constraints of Sec.~\ref{sec_lgas_fc}. Note that the
saturation curve (SC) (which is also called ``dew-point line'') and the
boiling curve (BC) (which is also called ``bubble-point line'') coincide in the
case of congruent PTs or azeotropic compositions, and are
split into separate boundaries in the general case of NCPT. The critical point
(CP) marked by the black dot, which is also a (critical) endpoint here, is
located at a
temperature of 14.75~MeV and baryon chemical potential of 912.4~MeV
(further values are given in Table~\ref{tab:cps}). It is known from
other studies, that the CP of LGS is usually also
the global maximum of the phase transition temperature, i.e.\ for all possible
values of $Y_Q$.

\begin{figure}[t]
\begin{center}
\includegraphics[width=\columnwidth]{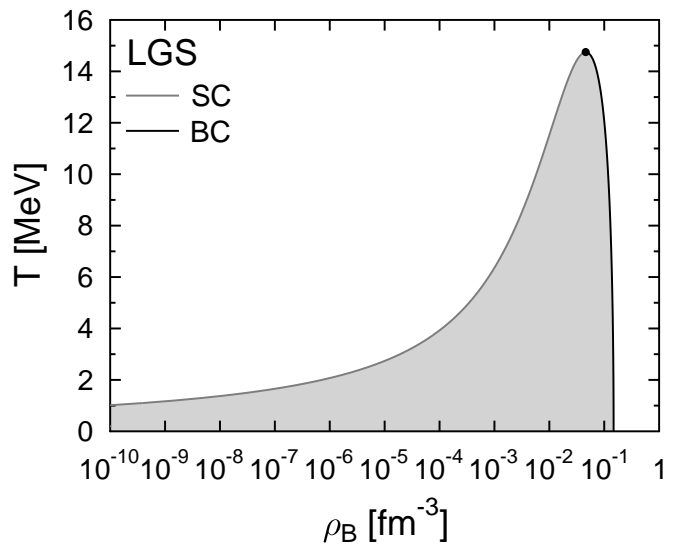}
\caption{\label{fig:tnb0.5}The binodal line which encloses the coexistence
region (filled with gray) in the $T$ - $\rho_B$ plane for case LGS. 
The black dot marks the critical point (CP). The
gray line to the left of the CP is the saturation curve (SC), and the black line
to the right is the boiling curve (BC).}
\end{center}
\end{figure}

In Fig.~\ref{fig:pt_lg_s} we show the pressure-temperature phase diagram,
where we also obtain a phase transition line. Note that the pressure on the
coexistence line goes to
zero in the zero-temperature limit. 
For a congruent PT the Clapeyron-equation is valid:
\begin{equation}
 \frac{dP}{dT}=\frac{s^I-s^{II}}{1/\rho_B^I-1/\rho_B^{II}} \; ,
\label{clapeyron}
\end{equation}
with $s^I=S^I/B^I$ and $s^{II}=S^{II}/B^{II}$ denoting the entropy per baryon
of the two phases. The Clapeyron-equation describes how the slope of the 
pressure-temperature phase transition line
is related to the difference in baryon number density and entropy per baryon of
the two phases. In our investigation, we always have $\rho_B^I<\rho_B^{II}$,
i.e. the first phase is assumed to have lower density. In Fig.~\ref{fig:pt_lg_s}
we see that $dP/dT>0$, and thus $s^G > s^L$ (where we have replaced ``$I$'' by ``$G$''
and ``$II$'' by ``$L$''). The gas phase has a higher entropy per baryon and is always
less dense than the liquid phase, which is a characteristic of the LGPT.

In Fig.~\ref{fig:tnb0.5} the binodal region is shown in the temperature-density
plane. The gray line to the left of the critical point
depicts the SC, where droplets of liquid form within the nucleon gas. The black
line is the BC, where bubbles of gas form inside the liquid. The region
enclosed by the two lines is the phase coexistence region, where a
two-phase mixture of gas and liquid is present. Here, and also 
in all following plots, filled areas correspond to states of
such a
two-phase coexistence. Due to the congruent behavior of LGS, for each point
inside the binodal or phase coexistence region, the gas state on the SC is in
coexistence with the liquid state on the BC at the same common temperature. Thus
the gas and the liquid are distinguished from each other by density, whereas the
liquid is always more dense. At the critical point the two
phases are identical. Inside the phase coexistence region, the volume fraction
of the liquid phase $\alpha= {V^L}/{V}$
and the gas phase ($1-\alpha$) are set by the total
baryon number density $\rho_B=B/V$ through:
\begin{equation}
\rho_B = \rho_B^G (1-\alpha)+ \rho_B^L \alpha  \; .
\end{equation}
Obviously, one has $\alpha = 0$ on the SC and $\alpha=1$ on the BC.

\begin{figure}[t]
\begin{center}
\includegraphics[width=\columnwidth]{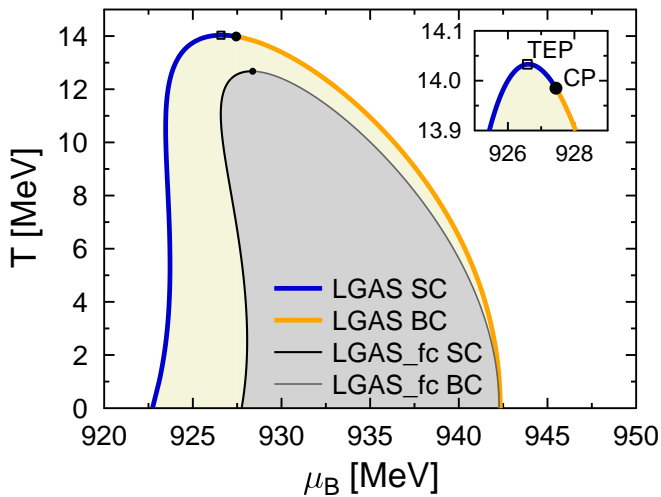}
\caption{\label{fig:tmub_lg_as}(Color online) Phase diagram in the temperature - baryon
chemical potential plane for the two asymmetric systems LGAS and LGAS\_fc with
$Y_Q = 0.3$. 
The gray and black thin lines show the
forced-congruent variant LGAS\_fc, and the corresponding small black dot is the
pseudo-critical point. The colored thick lines depict case LGAS, where $Y_Q$ is
not constrained locally. The corresponding black dot is the critical point
and the open square the temperature endpoint.
The inlay shows a zoom-in to the region around the critical point (CP) and 
temperature endpoint (TEP) of the
non-congruent case LGAS. The filled areas are the coexistence regions.}
\end{center}
\end{figure}
For asymmetric nuclear matter in the case LGAS one obtains a non-congruent phase
transition, which can be seen in Fig.~\ref{fig:tmub_lg_as}, depicted by the orange
and blue thick lines. For the non-trivial solution of the
equilibrium conditions in the non-congruent case we have used the method
described in Ref.~\cite{ducoin06}. The gray and black thin lines show the
forced congruent
variant
LGAS\_fc which will be discussed later. For LGAS one has a
phase coexistence region in $T$-$\mu_B$, enclosed by the orange and blue thick
lines, instead of a single line as in Fig.~\ref{fig:tmub_lg_s} for LGS.
Also in all following plots, we will
use colored thick lines for NCPTs. Thus one
can also distinguish between the branch belonging to the SC and the branch
belonging to the BC by different colors. 

As it was stressed in Refs.~\cite{iosilevskiy10,I5}, in non-congruent
VdW-like phase transitions of gas-liquid type there is no more a unique
``critical endpoint''. Instead,
three separate endpoints exist: maximal
temperature (cricondentherm)\footnote{The temperature endpoint
(TEP) or point of maximal temperature, which is also called the
``cricondentherm'' \cite{reid66,mcgraw}, is defined as the point with the
highest temperature where
phase coexistence is possible.}, 
maximal pressure (cricondenbar)\footnote{The pressure endpoint, also called ``cricondenbar''
\cite{reid66,mcgraw}, is
defined as the point on the binodal where the maximal pressure is obtained.},
and point of extremal chemical potential\footnote{The chemical potential endpoint
or point of extremal chemical
potential is defined as the point where the chemical potential of the binodal
surface is extremal with respect to temperature.}. 
In NCPT, these three
``topological'' endpoints are separated from the singular thermodynamic
object---the true non-congruent critical point\footnote{The critical point is
defined as the point on the binodal surface
where the two phases are identical. Because it is located on the binodal, an
infinitesimal change of the state can lead to phase separation into two phases
which can be distinguished from each other by an order parameter.}. 
Note also that the critical point of a congruent phase
transition is
determined by:
\begin{equation}
 \left.\frac{\partial P}{\partial \rho_B}\right|_{T}=\left.\frac{\partial^2
P}{\partial \rho_B^2}\right|_{T}=0 \; . \label{cpc}
\end{equation}
In contrast, for a NCPT this criteria is not applicable,
and the critical point does not fulfill it in general.

\begin{figure}[t]
\begin{center}
\includegraphics[width=\columnwidth]{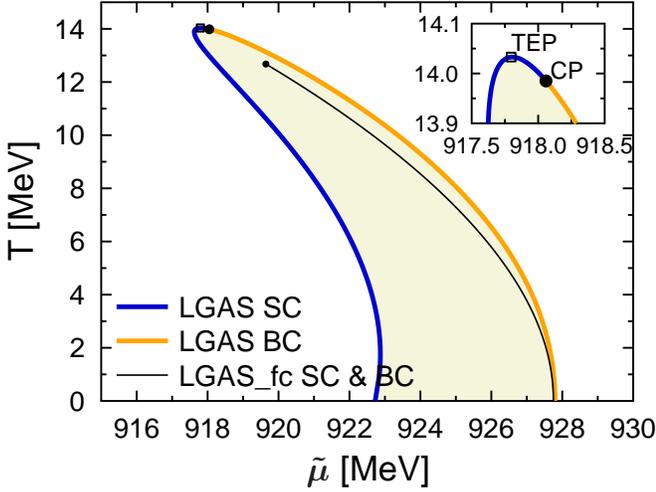}
\caption{\label{fig:tmutilde_lg_as}(Color online)
Phase diagram in the plane of temperature and
Gibbs free energy per baryon $\tilde \mu$ for case LGAS
and LGAS\_fc ($Y_Q=0.3$).
Otherwise the same notation and depiction as in Fig.~\ref{fig:tmub_lg_as} is
used.}
\end{center}
\end{figure}

\begin{figure}[t]
\begin{center}
\includegraphics[width=\columnwidth]{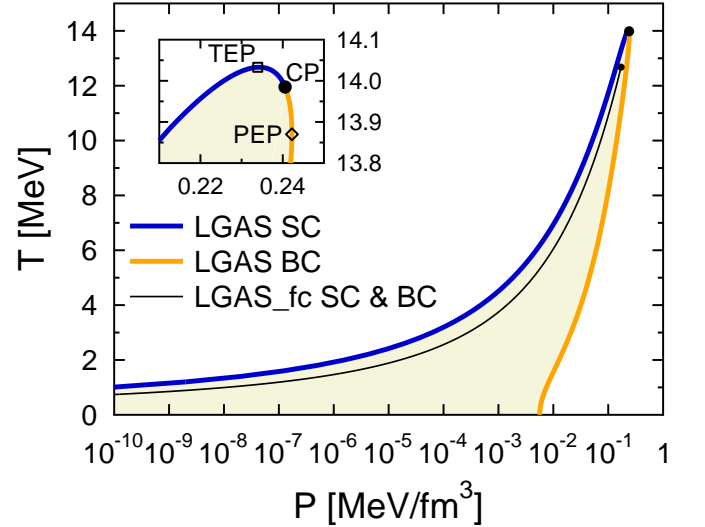}
\caption{\label{fig:pt_lg_as}(Color online) Phase diagram in the plane of temperature and 
pressure for cases LGAS and LGAS\_fc ($Y_Q=0.3$). In addition to the critical
point (black dot), the temperature (open square) and pressure
endpoints (open diamond) are shown in the inlay. Otherwise the same notation
and depiction as in Fig.~\ref{fig:tmub_lg_as} is used.}
\end{center}
\end{figure}

The inlay of Fig.~\ref{fig:tmub_lg_as} shows that the temperature
endpoint is
different from the critical
point. For LGAS, the
critical point is found at $T^{\rm CP} = 13.99$~MeV
(lower than in LGS) and $\mu_B^{\rm CP} = 927.4$~MeV. 
It is very interesting that this reduction of the critical temperature agrees 
very well with the experimental results of
Refs.~\cite{mcintosh12}.
The further
properties of the critical point are given in Table~\ref{tab:cps}. The temperature 
endpoint is located at
$T^{\rm TEP}=14.03$~MeV and $\mu_B^{\rm TEP}=926.6$~MeV. We remark that for
LGAS the
temperature
endpoint is located on the saturation curve (blue thick line), which in
principle could also be located on the boiling curve (orange thick line).
This
topology (i.e.,
location of the temperature endpoint on the two-phase boundary
relative to the critical point) 
is the same as for the gas-liquid NCPT in
uranium-oxygen plasma \cite{I1,I2,I3,I4,I5,iosilevskiy03}, which is taken as the
prototype of NCPT for the present study of LGAS (compare
Figs.~\ref{fig:tmutilde_lg_as} and \ref{fig:pt_lg_as} with Fig.~1 of
Ref.~\cite{iosilevskiy10}). 

In LGAS\_fc the two phases are constrained locally to have the same charge
fraction $Y_Q^G = Y_Q^{L}=0.3$. The results are depicted by the gray
and black thin lines
in Fig.~\ref{fig:tmub_lg_as}. The two lines also enclose a phase coexistence
region, which illustrates the non-equality of $\mu_B$ of the two phases in the
phase coexistence region, due to the locally constrained charge fraction (see
also the discussion in Sec.~\ref{sec_lgas_fc} and Appendix
\ref{app_mutot_lghifc}).
The Gibbs free energy per baryon $\tilde \mu$ is the only chemical potential
which is equal in the two phases. Furthermore, for isothermal phase
transitions it is a constant, because the properties of the two phases do
not change. In contrast to the non-congruent phase transition LGAS, for the
forced-congruent phase transition LGAS\_fc $\mu_B$ is dependent on the
baryon number density $\rho_B$ and given by:
\begin{equation}
 \mu_B = \mu_B^G \frac{\rho_B^G}{\rho_B}(1-\alpha) + \mu_B^L
\frac{\rho_B^L}{\rho_B}\alpha \; , \label{eq_mublgasfc}
\end{equation}
which is derived in Appendix \ref{app_mutot_lghifc}.

The phase diagrams as a function of the Gibbs free energy per baryon $\tilde \mu$ are shown in
Fig.~\ref{fig:tmutilde_lg_as} for both cases LGAS and LGAS\_fc. The
banana-shaped region of LGAS is typical for non-congruent liquid-gas like phase
transitions, see Refs.~\cite{I1,I2,I3,I4,I5,iosilevskiy03}.
In Fig.~\ref{fig:tmutilde_lg_as} the congruence of LGAS\_fc becomes obvious.
Comparing LGAS and LGAS\_fc, the phase coexistence region turns into a phase
coexistence line, when enforcing the local constraint for the charge fraction. 
Furthermore, for LGAS\_fc the pseudo-critical point (properties listed in Table
\ref{tab:cps}) coincides with the temperature and chemical potential endpoints.
Note that the pseudo-critical point of a forced-congruent phase transition obeys
Eq.~(\ref{cpc}). We remark that the phase transition line of the
forced-congruent variant
must lie strictly inside the two-phase region of the non-congruent phase
transition \cite{I1,I2,I3,I4,I5,iosilevskiy03}, which also can be seen as a
consequence of Le Chatelier's principle. As an exception, both objects could
touch each other in azeotropic points of the parameter space, as seen 
for LGS. 

Note that for the LGS case, $\mu_B=\tilde \mu$, since $\mu_Q\equiv0$. Thus the
phase-coexistence line of LGAS\_fc in Fig.~\ref{fig:tmutilde_lg_as} can be directly
compared with the one of LGS in Fig.~\ref{fig:tmub_lg_s} and it is found that
their shape is very similar. However, states on the phase coexistence line of
LGAS\_fc in Fig.~\ref{fig:tmutilde_lg_as} belong to two different values of the
baryon chemical potential, shown by the two gray and black thin lines in
Fig.~\ref{fig:tmub_lg_as}. If $\tilde \mu$ in LGAS\_fc were changed in a
continuous way and the phase transition line in Fig.~\ref{fig:tmutilde_lg_as} is
crossed, $\mu_B$ jumps from the value of the gas phase to the value of the
liquid phase in Fig.~\ref{fig:tmub_lg_as}. This can be seen as a sign of the
\textit{enforced} congruence, in contrast to the azeotropic congruent phase
transition LGS.

Figure \ref{fig:pt_lg_as} shows the same scenarios as the previous figures,
but gives the phase diagrams
in the temperature-pressure plane. Here it is clear, that the phase
transition in the forced congruent variant LGAS\_fc occurs only at a single
value of the pressure, which is the same behavior as in Fig.~\ref{fig:pt_lg_s} for LGS. In contrast,
for a given temperature in LGAS, there is an extended coexistence range in
pressure, which is enclosed by the SC and BC. The names ``SC'' and ``BC'' 
are widely accepted for non-congruent evaporation in chemically reacting 
plasmas \cite{I1,I2,I3,I4,I5} where there meaning is obvious. They are also most intuitive for LGAS
for this kind of phase diagram: for a fixed pressure of e.g.\
$10^{-2}$~MeV/fm$^{3}$ and starting from $T=0$, by heating the
system one will reach the boiling curve, where bubbles of gas
appear inside the liquid. Conversely, if one starts at high temperatures and cools
the system isobarically, droplets of liquid will form within the gas when the
saturation curve is reached. In this figure we can also identify the pressure
endpoint which is located on the BC of LGAS. 
In the gas-liquid NCPT
in uranium-oxygen plasma \cite{I1,I2,I3,I4,I5,iosilevskiy03}, which is the
prototype for our present study of NCPT in LGAS, one has the same topology 
that the pressure endpoint is located on the BC, despite differences
in the thermodynamic variables by many orders of magnitude (compare 
Fig.~\ref{fig:pt_lg_as}
with Fig.~1 in Ref.~\cite{iosilevskiy10}). For LGAS\_fc, all three endpoints
coincide with the critical point defined by Eq.~(\ref{cpc}) (see Fig.~1 in
Ref.~\cite{iosilevskiy10}).

\begin{figure}
\begin{center}
\includegraphics[width=\columnwidth]{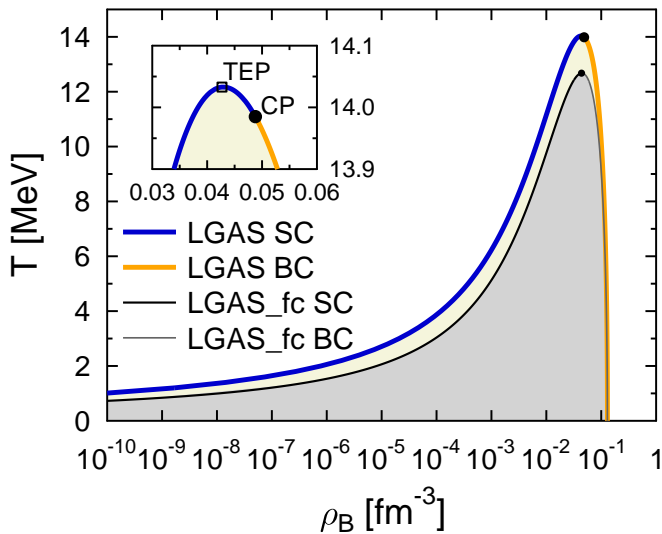}
\caption{\label{fig:tnb0.3}(Color online) The binodal line enclosing the coexistence region in
the $T$ - $\rho_B$ plane for LGAS and LGAS\_fc ($Y_Q=0.3$). Otherwise the same notation and
depiction as in Fig.~\ref{fig:tmub_lg_as} is used.}
\end{center}
\end{figure}
In spite of the similarity of the NCPT in LGAS with its
terrestrial prototypes \cite{I1,I2,I3,I4,I5,iosilevskiy03}, the
significant difference in the topology of P--T diagrams should be
stressed (compare Fig.~\ref{fig:pt_lg_as} above with Fig.~1 in 
Ref.~\cite{iosilevskiy10}). While the
pressure on both boundaries of the non-congruent PT in the uranium-oxygen system
\cite{I1,I2,I3,I4,I5,iosilevskiy03}---boiling and saturation curves---tends to
zero for the limit $T\rightarrow 0$, the pressure on the boiling curve in NCPT
in asymmetric ($Y_Q=0.3$) LGPT does not tend to zero when $T\rightarrow 0$
in our case (see
Fig.~\ref{fig:pt_lg_as}). The same feature was noted already for the NCPT of 
asymmetric nuclear matter calculated with a different EOS (Fig.~3 in Ref.~\cite{maruyama10}). 
The
reason
for this feature is the difference in the physical nature of the involved
forces which are relevant for the non-congruence in a chemically-reacting
uranium-oxygen plasma \cite{I1,I2,I3,I4,I5,iosilevskiy03} and in asymmetric
nuclear matter, and also the use of Fermi-Dirac statistics for the latter.

In Fig.~\ref{fig:tnb0.3} we show the binodal or phase coexistence regions for
LGAS and LGAS\_fc in the temperature - density plane, similar as in
Fig.~\ref{fig:tnb0.5} for LGS. Again, the non-congruent behavior of LGAS can be
identified by the non-equivalence of the temperature endpoint and the critical
point. Conversely, for LGAS\_fc the endpoints and the (pseudo-) critical point
coincide. Furthermore, for isothermal processes of LGAS at  temperatures $T^{\rm
CP}<T<T^{\rm TEP}$ so-called retrograde condensation occurs (see also
Ref.~\cite{barranco80,muller95}): imagine, e.g., an isothermal compression at
$T=14$~MeV. First one hits the saturation curve from the left, and a liquid with
a larger $Y_Q$ and a larger $\rho_B$ appears inside the gas phase. With
increasing density the volume fraction of the liquid will first increase. But
for retrograde condensation, for densities larger than a certain density, the
volume fraction will decrease again, until it returns to zero at the right side
of the saturation curve. The liquid has disappeared again after the phase
coexistence region has been crossed.

\begin{figure}
\begin{center}
\includegraphics[width=\columnwidth]{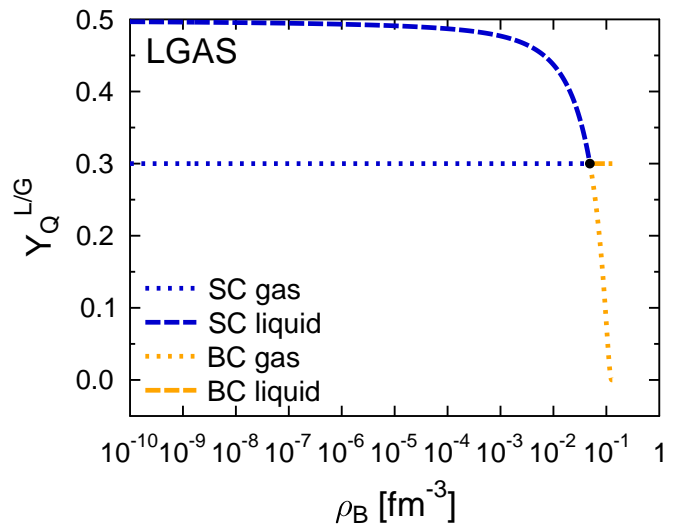}
\caption{\label{fig:ynb0.3}(Color online) The charge fraction of the liquid (dashed lines) and
the gas phase (dotted lines) for case LGAS as a function of the coexistence 
density $\rho_B$ along
the binodal line of Fig.~\ref{fig:tnb0.3}. The blue lines show the charge fractions 
of the two
coexisting phases for states on
the saturation curve presented in previous figures, the orange lines for states 
on the boiling curve.}
\end{center}
\end{figure}
The non-congruent behavior of LGAS is further analyzed in the following plots
of local ``chemical'' composition and density.
Fig.~\ref{fig:ynb0.3} shows the charge fractions of the two phases which are in
coexistence, if one moves along the liquid and vapor binodal lines of
Fig.~\ref{fig:tnb0.3}. The blue dotted line ``SC gas'' in Fig.~\ref{fig:ynb0.3}
depicts the charge 
fraction of the gas phase
$Y_Q^G$ for the states on the saturation curve shown in Fig.~\ref{fig:tnb0.3}. 
On the saturation curve, the gas phase is in coexistence with a liquid phase 
which has a different charge fraction $Y_Q^L$, shown by the dashed blue line.
Also for the conditions of the boiling curve of Fig.~\ref{fig:tnb0.3}, we 
always have coexistence of a gas phase (orange dotted line in
Fig.~\ref{fig:ynb0.3}) with a liquid phase 
(orange dashed line in
Fig.~\ref{fig:ynb0.3}), which have different charge fractions.

\begin{figure}
\begin{center}
\includegraphics[width=\columnwidth]{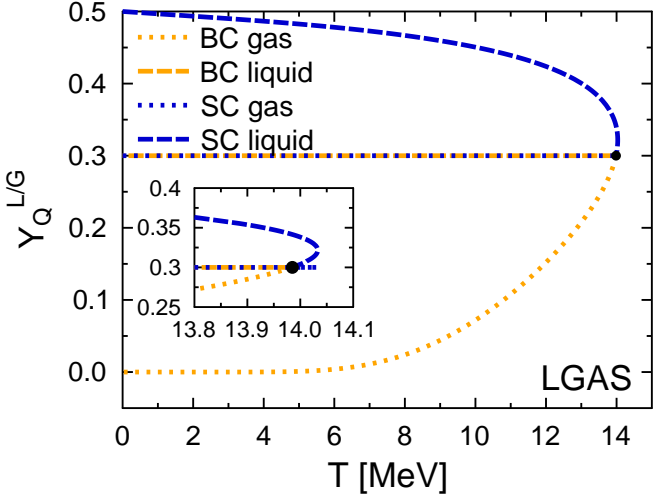}
\caption{\label{fig:yt0.3}(Color online) As Fig.~\ref{fig:ynb0.3}, but now in dependence of
the coexistence temperature of the binodal line of Fig.~\ref{fig:tnb0.3}.
The inlay shows a zoom-in to the region around the critical point, including the region of so-called retrograde condensation at $T>T^{\rm CP}$.}
\end{center}
\end{figure}

\begin{figure}
\begin{center}
\includegraphics[width=\columnwidth]{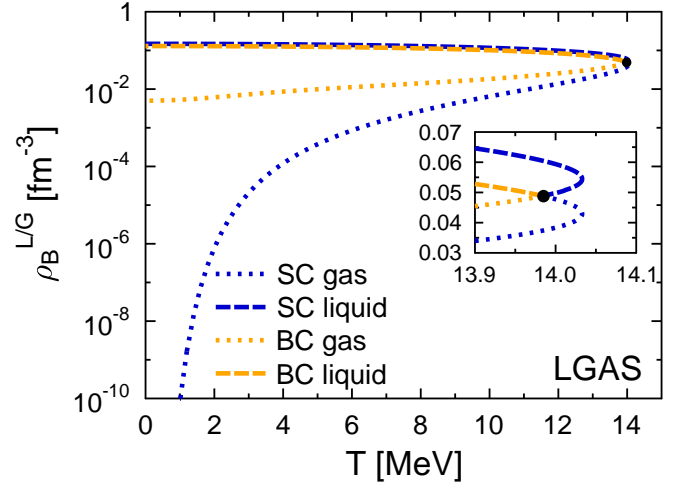}
\caption{\label{fig:nbt0.3}(Color online) The baryon number density of the liquid (dashed
lines) and the gas phase (dotted lines) for case LGAS as a function of the coexistence
temperature on the binodal line of Fig.~\ref{fig:tnb0.3}. Otherwise the same
notation and depiction as in Fig.~\ref{fig:yt0.3} is used.}
\end{center}
\end{figure}

In
all previous plots, the depicted quantities correspond to total thermodynamic
quantities, i.e.\ of the system as a whole. In contrast, in
Figs.~\ref{fig:ynb0.3}, \ref{fig:yt0.3}, and \ref{fig:nbt0.3}
we are showing individual properties of the two coexisting phases. Now and
in the following
we are using dashed and dotted lines in such plots to illustrate this
difference. The color coding helps to identify the same states in the different
diagrams. For example in Fig.~\ref{fig:ynb0.3}, the ends of the orange
curves,
which correspond to $T=0$, are
given by the highest density of coexistence of LGAS in Fig.~\ref{fig:tnb0.3}
which is on the orange solid line.

In Fig.~\ref{fig:ynb0.3}, one of the two phases always must have
$Y_Q=0.3$, whereas the charge fraction of the second phase is not constrained,
because its volume fraction is still zero on the binodal line. For states on the
saturation curve one is still in the gas phase, i.e.\ $\alpha=0$, thus
$Y_Q^G=0.3$. For states on the boiling curve one is still in the liquid phase,
$\alpha = 1$, and $Y_Q^L=0.3$. The charge fraction is an order parameter for
LGAS and thus it can be used to characterize the two phases, with the
identification that the gas phase always has a lower charge fraction than the
liquid, i.e.\ $Y_Q^G < Y_Q^{L}$. Only at the critical point is equality
established, $Y_Q^G = Y_Q^{L}$. It is interesting to note, that the charge
fraction of the phase with $Y_Q \neq 0.3$ shows the same dependence on density
before and after the critical point.

In Fig.~\ref{fig:yt0.3} we also show the charge fractions of the two phases
along the binodal line, but now as a function of the coexistence temperature. By
comparing with Figs.~\ref{fig:tmub_lg_as} - \ref{fig:tnb0.3}, it is obvious,
that for each coexistence temperature there are always two points on the binodal
line, corresponding to two different halves of the binodal line which are
separated from each other by the temperature endpoint. For each half, two phases
with different values of $Y_Q$ are in coexistence. Consequently, in
Fig.~\ref{fig:yt0.3}, for each temperature there are always four values of
$Y_Q$.
For $T< T^{\rm CP}$, one has $Y_Q^G=0.3$ and $Y_Q^L > 0.3$ on the saturation
curve, and $Y_Q^G<0.3$ and $Y_Q^L = 0.3$ on the boiling curve. Note that the two
lines with $Y_Q=0.3$ are on top of each other. For $T^{\rm CP}<T<T^{\rm TEP}$,
both halves belong to the saturation curve, and thus $Y_Q^G=0.3$ for both
halves, each being in coexistence with a different configuration of the liquid
with $Y_Q^L>0.3$. 

The previous two figures can be used to identify the high degree of isospin
distillation of LGAS in the
limit $T \rightarrow 0$. Let
us consider a decompression at $T \sim 0$ of the asymmetric system with
$Y_Q=0.3$. Once the boiling curve is reached vapor bubbles appear which in this
case consist of pure neutron gas, $Y_Q^G \rightarrow 0$. Obviously, this leads to the
distillation of a symmetric liquid by evaporation of pure neutron
bubbles from a boiling asymmetric liquid phase. On the other hand, for
saturation conditions (``dewpoint'') at $T \sim 0$, liquid microdrops
tend to the exactly symmetric composition, $Y_Q^L \rightarrow 0.5$.  These
features of the NCPT of case LGAS differ significantly from the behavior of the
chemical composition (O/U-ratio) in NCPTs of uranium-oxygen systems (compare
Fig.~\ref{fig:yt0.3} with Fig.~2 of Ref.~\cite{I1} and  Fig.~3 of
Ref.~\cite{I2}).

In a similar way as in Fig.~\ref{fig:yt0.3}, in Fig.~\ref{fig:nbt0.3} we show
the baryon number densities of the two phases for each of the two halves of the
binodal line as a function of the coexistence temperature.
Presented in this way, one sees that the density is also an order parameter of
LGAS, whereas the liquid is the phase with the higher
density. At the critical point, the liquid and the gas have the same density and
charge fraction and thus cannot be distinguished from each other any longer. The
discussion of this figure is similar as for Fig.~\ref{fig:yt0.3}: For $T< T^{\rm
CP}$ there are the $\rho_B$-curves from the liquid (blue dashed line) and the gas (blue
dotted line) on the saturation curve, and another pair of $\rho_B$-curves from the
liquid (orange dashed line) and the gas (orange dotted line) on the boiling curve.
For $T^{\rm CP}<T<T^{\rm TEP}$ there are still the four different values of
$\rho_B$. However, now all points belong to the saturation curve.

\subsection{Deconfinement phase transition}
\begin{figure}
\begin{center}
\includegraphics[width=0.9\columnwidth]{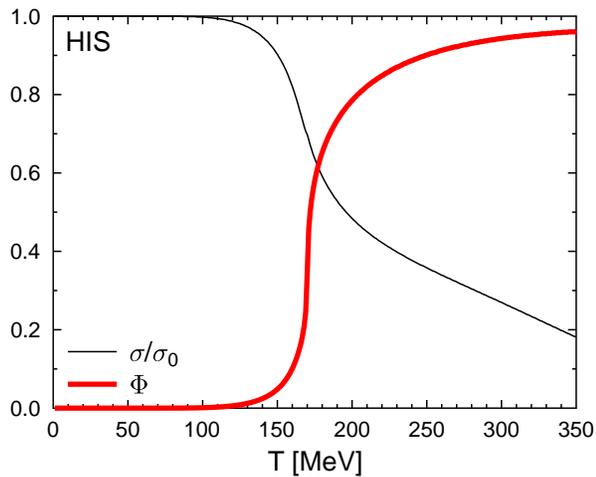}
\caption{(Color online) Order parameter for chiral symmetry
restoration/breaking $\sigma$ normalized by the vacuum value $\sigma_0$ and order parameter for
deconfinement/confinement $\Phi$ versus temperature at zero baryon chemical
potential for symmetric matter of case HIS ($Y_Q=0.5$).}
\label{pol}
\end{center}
\end{figure}
\begin{figure}
\begin{center}
\includegraphics[width=\columnwidth]{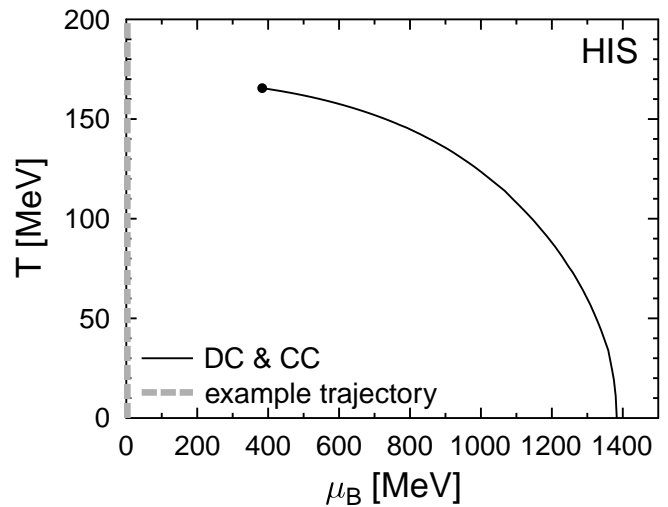}
\caption{\label{fig:05TxMu}Phase diagram in the temperature - baryon chemical
potential plane for case HIS  ($Y_Q=0.5$), for which the deconfinement curve (DC) coincides with the
confinement curve (CC). The black dot marks the critical point. The thick gray
line along the y-axis shows part of the example trajectory belonging to
Fig.~\ref{pol}.}
\end{center}
\end{figure}

For the LGPT we used the baryon number density and charge fraction as order
parameters. For the deconfinement and chiral symmetry PTs,
typically the Polyakov-loop $\Phi$ and the chiral condensate $\sigma$ are used,
as already discussed in Sec.~\ref{sec_su3}. The field $\sigma$ 
characterizes chiral
symmetry restoration whereas $\Phi$ can be taken as a measure for deconfinement.
At finite
temperature and $\mu_B=0$, respectively $\rho_B=0$,\footnote{We remind
the reader that we include anti-particles, therefore if $T>0$, we have
equivalence between $\rho_B=0$ and $\mu_B=0$. Furthermore, for $T>0$, $\mu_B<0$
corresponds to net anti-matter with $\rho_B<0$, which is not relevant here.
For $T=0$ the situation is a bit more complicated, because of the LGPT which
extends down
to $\rho_B=0$ at a constant finite value of $\mu_B$.} there is no first-order PT,
but a smooth crossover between the hadronic (confined, chiral symmetry broken)
and the quark phase (deconfined, chiral symmetry partly restored)
\cite{aoki06}. This is shown in Fig.~\ref{pol} for HIS, where the ratio
$\sigma/\sigma_0$ decreases from
one to lower values and
$\Phi$ goes from zero to a value close to one in a smooth fashion. We
remark that we define the
cross-over temperature $T^{\rm co}$ as the peak of the change of the chiral
condensate and $\Phi$ with $T$, yielding a value of $T^{\rm co}= 171$~MeV,
in accordance with lattice QCD results \cite{Fodor:2004nz}. 
This behavior of the order parameters
corresponds to the example trajectory through the phase diagram of
HIS shown in Fig.~\ref{fig:05TxMu}.

For high enough baryon number densities, the QHPT turns into a first-order phase
transition. This can be seen in Fig.~\ref{fig:05TxMu} where we show the
first-order phase transition line for case HIS, i.e.\ for heavy-ion collisions
of symmetric nuclei. The critical point is located at $T^{\rm CP}=165.5$~MeV and
$\mu_B^{\rm CP}=383$~MeV, again, in accordance with lattice QCD results 
\cite{Fodor:2004nz}.
Its further properties are listed in
Table~\ref{tab:cps}. The topology of this PT is the same as in
LGS, see Fig.~\ref{fig:tmub_lg_s}, only the typical scales are
different. For
example, the critical point of HIS is at a roughly ten times higher temperature.
For the QHPT, we will use the terms ``deconfinement curve'' (DC) instead of
``saturation curve'' and ``confinement
curve'' (CC) instead of ``boiling curve'', which we think is more meaningful.
If coming from low densities and temperatures, first droplets of
denser deconfined
quark matter will appear when the DC is reached. Conversely, if coming from high
densities and temperatures, when the CC is reached, the first quarks will start
to be confined into bubbles of less dense hadronic matter. 
There is an interesting analogy to the ``ionization boundary curve'' and 
``recombination boundary curve'' of the hypothetical ionization-driven plasma 
phase transition in dense hydrogen at megabar pressure range (see, e.g., \cite{fortov2006,fortov2011,iosi_eltp2000}). This first-order PT in 
weakly ionized hydrogen (predominantly H and H$_2$ and small amount of p and e)
is driven by a jump-like ionization (deconfinment) into 
highly ionized hydrogen (predominantly p and e and small amount of H and H$_2$).
We want to point out the similarity between the hydrogen plasma which is an 
arbitrary solution of H$_2$, H, p, and e, and the Chiral SU(3) model in which 
quarks and hadrons can in principle also be mixed in arbitrary proportions 
(nevertheless a clear distinction of the two phases is always possible, see Sec.~\ref{sec_su3}). 

HIS is, in principle, a binary
system, with baryon number and electric charge (respectively isospin) as two
globally conserved charges (see Table~\ref{tab:cases}). But the PT in HIS is azeotropic,
meaning it is congruent and the Maxwell construction can be used, just like symmetric
nucleonic matter in LGS.
It is not so
obvious as for LGS that matter in HIS is an azeotrope, because a whole set of particles,
including strange ones, is considered (see Table~\ref{tab:cases}). However,
strange quarks and hyperons do not invalidate the relation
between isospin symmetry and azeotropic behavior, if strangeness is set locally
to zero, as it is done here. In this case, the total density of strange quarks,
i.e.\ in form of unbound strange quarks or bound in hyperons, is equal to the
total density of anti-strange quarks. A more detailed explanation of the
matter is given in Appendix \ref{app_azeo}.

\begin{figure}[t]
\begin{center}
\includegraphics[width=\columnwidth]{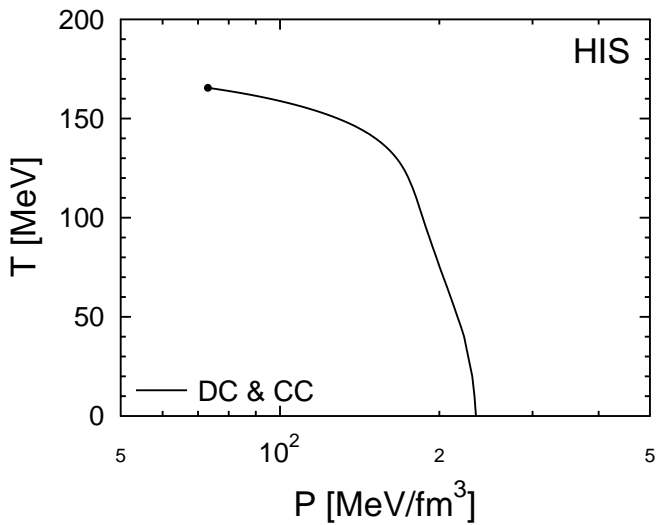}
\caption{\label{fig:05Txp}Phase diagram in the temperature - pressure plane for
case HIS. Otherwise notation and depiction as in Fig.~\ref{fig:05TxMu}.}
\end{center}
\end{figure}
\begin{figure}[t]
\begin{center}
\includegraphics[width=\columnwidth]{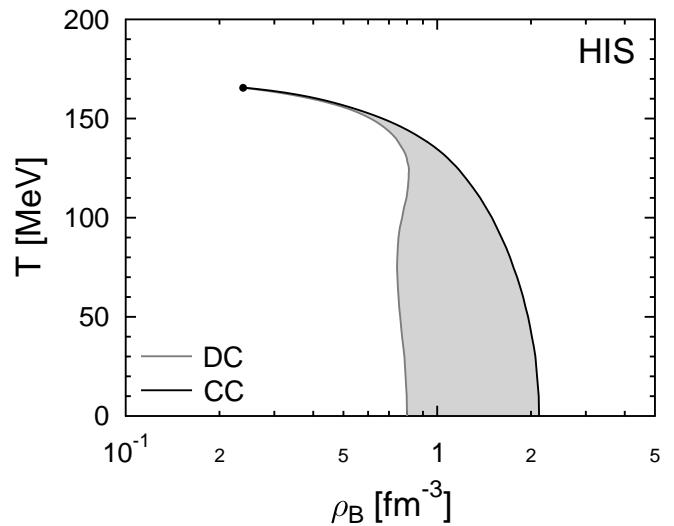}
\caption{\label{fig:05Txrho}Phase diagram in the temperature - baryon number
density plane for case HIS. The deconfinement curve (DC) is to the left of the
confinement curve (CC). The gray area shows the two-phase coexistience region.
Otherwise notation and depiction as in
Fig.~\ref{fig:05TxMu}.}
\end{center}
\end{figure}

In Fig.~\ref{fig:05Txp} the phase diagram in the pressure-temperature plane is
depicted. Comparing with Fig.~\ref{fig:pt_lg_s} one realizes an important
difference between the QHPT and the LGPT: the slope of the phase transition line
is negative. Therefore, the QHPT is not of liquid-gas type. With the Clapeyron
equation (\ref{clapeyron}) one finds that this is due to the fact that the
hadronic phase, which is less dense, has a lower entropy per baryon than the
quark phase, $s^H < s^Q$ (where we have replaced ``$I$'' by ``$H$'' and ``$II$''
by
``$Q$''), which is opposite to the behavior in the LGPT. The negative slope of
the
$p-T$ phase diagram makes the QHPT fundamentally different from the LGPT.
This fact (negative slope of the $p-T$ boundary for QHPT in a symmetric
system) is not absolutely new (e.g., presentations of I.I.\ at several conferences\footnote{See, e.g., http://theor.jinr.ru/\midtilde cpod/Talks/240810/Iosilevskiy.pdf.} and discussion in Ref.~\cite{iosi_eltp2000}, L.~Satarov, private communication
(2010) based on calculations via EOS model described in \cite{satarov09}, J.~Randrup, presentations at several conferences\footnote{See, e.g., http://theor.jinr.ru/\midtilde cpod/Talks/260810/Randrup.pdf} and Ref.~\cite{randrup12}, or \cite{yudin13}) but is
not well-recognized yet. Further investigation and analysis of
this fundamental difference between LGPT and QHPT is in preparation
\cite{iosilevskiy12}. Fig.~\ref{fig:05Txrho} shows the coexistence region in the temperature -
baryon number density plane, in a similar way to Fig.~\ref{fig:tnb0.5} for
the LGPT. Note that the shape of the phase coexistence region of HIS is
rather different from LGS. This is once more a manifestation (and not the last
one) of
the fundamental difference between LGPT and QHPT. 

\begin{figure}
\begin{center}
\includegraphics[width=\columnwidth]{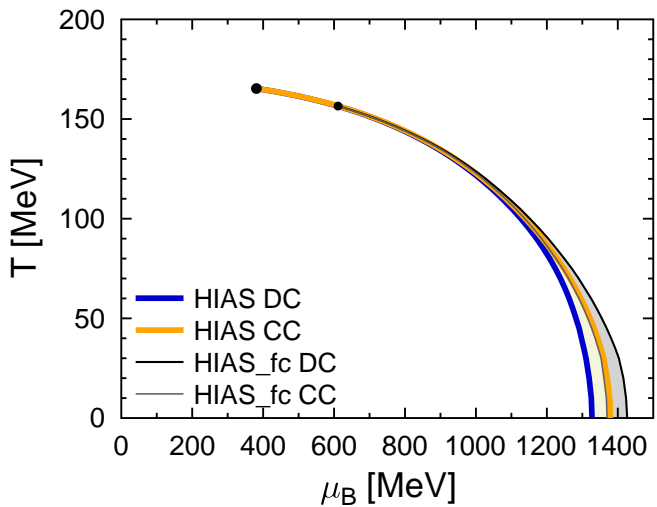}
\caption{\label{fig:03TxMu}(Color online) Phase diagram in the temperature - baryon chemical
potential plane for cases HIAS and HIAS\_fc ($Y_Q=0.3$). The colored thick lines
belong to HIAS, the gray and black thin ones to HIAS\_fc. 
The large black dot marks the
critical point of HIAS and the small black dot the pseudo-critical point
of HIAS\_fc. The filled
areas show the coexistence regions. Note that for HIAS\_fc the confinement
curve (CC) is to the left of the deconfinement curve (DC).}
\end{center}
\end{figure}
In case HIAS of heavy ion collisions with asymmetric Coulomb-less matter
one has a true
binary system. The PT is
non-congruent and the Gibbs construction must be used. This is visible in
Fig.~\ref{fig:03TxMu}, where one obtains a phase-coexistence region instead of a
phase-transition line as in HIS before. We can compare HIAS with LGAS of
Fig.~\ref{fig:tmub_lg_as}. Obviously, the phase-coexistence region is much
narrower than for LGAS, if we compare the width in $\mu_B$ relative to the
extension in temperature. HIAS\_fc is the forced congruent variant, where the
charge fraction is constrained locally, $Y_Q^H=Y_Q^Q=0.3$, so that the Maxwell
construction can be used. The gray thin line is the corresponding DC, and the
black thin line the CC, which is partly covering the CC of HIAS. Very
interestingly, the DC and CC changed order for HIAS\_fc compared to HIAS. We
will explain this interesting result in detail later.

\begin{figure}
\begin{center}
\includegraphics[width=\columnwidth]{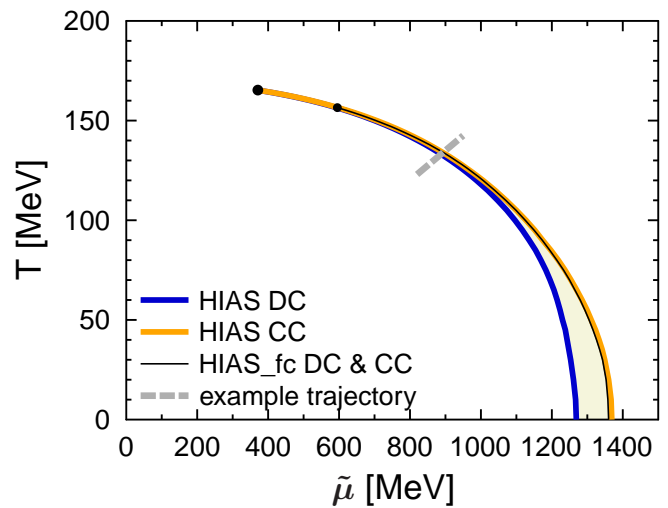}
\caption{\label{fig:03TxMuhat}(Color online) Phase diagram in the plane of temperature and 
Gibbs free energy per baryon $\tilde \mu$ for cases HIAS
and HIAS\_fc ($Y_Q=0.3$). The
thick gray line gives an example for a trajectory through the phase
diagram for which we show the order parameters in Fig.~\ref{fig:03PolxMuhat}.
Otherwise the same depiction and notation as in Fig.~\ref{fig:03TxMu} is used.
Note that the phase transition line of HIAS\_fc is very close to the
confinement curve (CC) of HIAS.}
\end{center}
\end{figure}
\begin{figure}[t]
\begin{center}
\includegraphics[width=\columnwidth]{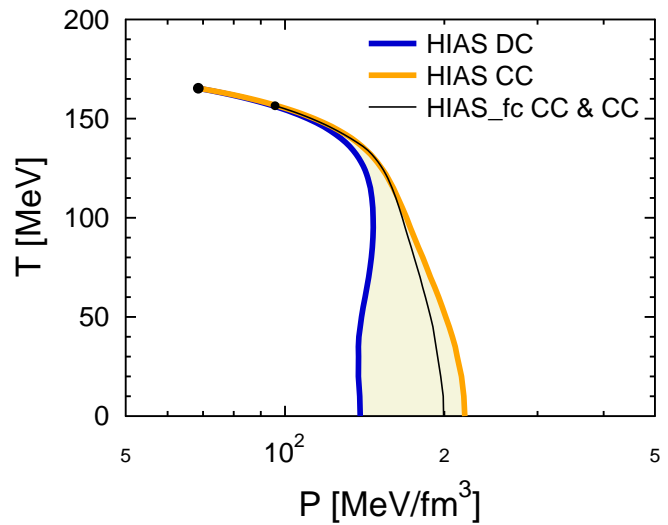}
\caption{\label{fig:03Txp}(Color online) Phase diagram in the
temperature - pressure plane for HIAS and HIAS\_fc ($Y_Q=0.3$). The same depiction
as in Fig.~\ref{fig:03TxMu} is used.}
\end{center}
\end{figure}

The phase diagrams as a function of the Gibbs free energy per baryon $\tilde
\mu$
are shown in Fig.~\ref{fig:03TxMuhat}, and as a function of pressure in
Fig.~\ref{fig:03Txp}. Here one sees that HIAS\_fc is a
congruent PT, because the DC and CC are identical. Furthermore,
the phase transition line of HIAS\_fc is inside the phase coexistence region of
HIAS, as it has to be. Comparing Fig.~\ref{fig:03TxMuhat} with
Fig.~\ref{fig:tmutilde_lg_as} of LGAS, again
one sees that the shape of the
coexistence region of HIAS is much narrower. This could be described as a weaker
non-congruence of the phase transition HIAS.

Note that in our calculations for HIAS we could not resolve the differences
between the temperature and pressure endpoints and the critical point. In
principle, around
the critical point a similar structure as for LGAS has to occur. We predict that
the phase coexistence regions of HIAS around the critical point are smooth and
two-dimensional, like for LGAS in the inlays of
Figs.~\ref{fig:tmub_lg_as}-\ref{fig:tnb0.3}, whereas the temperature endpoint
and critical point could also have their orders inversed. To be more precise, we expect a
round, ``banana-shaped'' region (see, e.g., Fig.~1 in
\cite{iosilevskiy10}). However, due to the fact that for lower
chemical potentials or baryon densities the first-order phase transitions HIS and HIAS
become rather weak, i.e., the phases on both sides of the transition become
extremely similar, this structure is not as easily observed anymore as it is
in the LGPT of nuclear matter (shown in the previous subsection) and in
chemically reacting plasmas
\cite{I1,I2,I3,I4,I5,iosilevskiy03}. The highest
temperature, for which we still could solve the equilibrium conditions was at
$T=165.3$~MeV for $\mu_B=381$~MeV, respectively $\tilde \mu = 371$~MeV where we
used a resolution of roughly 
$0.1$~MeV in 
temperature. Because this resolution cannot be resolved on the scale of
Fig.~\ref{fig:03TxMu}, we take these values of $T$, $\mu_B$, and $\tilde \mu$ as
the approximate temperature and chemical potentials of the critical point,
which are also listed in Table~\ref{tab:cps}. The very narrow phase
coexistence region shows that the two phases are in an extremely similar
thermodynamic state. For even higher temperatures, the
differences between the two
phases become smaller than our numerical accuracy, preventing us to find a more
precise solution. 

\begin{figure}[t]
\begin{center}
\includegraphics[width=\columnwidth]{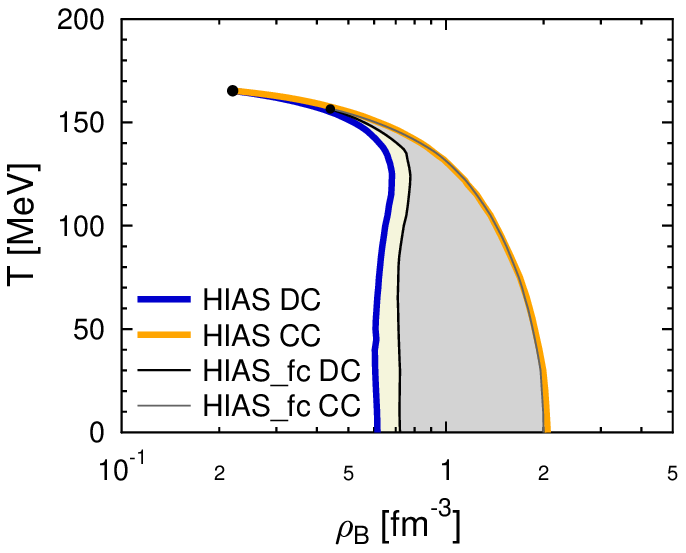}
\caption{\label{fig:03TxRho}(Color online) Phase diagram in the plane of temperature and baryon
number density for cases HIAS and HIAS\_fc ($Y_Q=0.3$). Otherwise the same depiction and
notation as in Fig.~\ref{fig:03TxMu} is used. The deconfinement curve (DC)
of HIAS\_fc is now to the left of the confinement curve (CC).}
\end{center}
\end{figure}

This aspect is also very pronounced in Fig.~\ref{fig:03TxRho} where the phase
diagrams of HIAS and HIAS\_fc are shown as a function of baryon number
density. At $T=0$, the DC and the CC have very different densities. Conversely,
for increasing temperatures the extension of the coexistence region in density
becomes extremely narrow. Note that the coexistence region of HIAS\_fc is within
HIAS, as it must be. 

There is one further important aspect. If we compare the phase diagram of
the asymmetric system, e.g.~Fig.~\ref{fig:03TxMu}, with the symmetric system in
Fig.~\ref{fig:05TxMu}, and the corresponding numbers in Table~\ref{tab:cps},
one finds that the critical points are
practically unaffected by the change of the asymmetry. $T$ and $\mu_B$ of HIS and
HIAS differ by less than 0.5~\%. In LGPT, a stronger dependence on the asymmetry
is observed. For LGS and LGAS the change from $Y_Q=0.5$ to $Y_Q=0.3$ leads to a
significant shift of the critical point, e.g. on the order of 6 \% in
temperature. This is naturally explained by the high temperatures of order
160~MeV around the critical point of HI. The effect of the asymmetry becomes
extremely weak, because the EOS in this regime is dominated by thermal
contributions. Obviously, for $\mu_B\longrightarrow 0$ the EOS would be
completely independent of the asymmetry. We conclude that even though HIAS is in
principle a non-congruent PT, in our calculations with the Chiral SU(3) EOS
model this non-congruence is almost negligible close to
the critical point. This is also visible in Figs.~\ref{fig:03TxMu} and
\ref{fig:03TxMuhat}, where the width of the coexistence region is smaller than
the line thickness of the curves.

Let us now come back to the explanation of the inverted ordering of DC and
CC in
HIAS\_fc in Fig.~\ref{fig:03TxMu}. For an isothermal compression, corresponding
to a horizontal line through Fig.~\ref{fig:03TxRho}, $\tilde \mu$ and
$\mu_B$ will increase
monotonically, until the deconfinement curve is reached at a certain density
$\rho_B=\rho_B^H$. This state corresponds to the value of $\mu_B=\mu_B^H$ in the
hadronic phase on the deconfinement curve, i.e.\ the black thin line most to the right
in Fig.~\ref{fig:03TxMu}.  For an isothermal compression, the corresponding
value of $\tilde \mu$ will remain constant throughout the phase transformation,
and $\tilde \mu = \tilde \mu^H = \tilde \mu^Q$. However, inside the coexistence
region $\mu_B$ will change with density because it is given by:
\begin{equation}
 \mu_B = \mu_B^H \frac{\rho_B^H}{\rho_B}(1-\alpha) + \mu_B^Q
\frac{\rho_B^Q}{\rho_B}\alpha \; , \label{eq_mubhiasfc}
\end{equation}
with the volume fraction of the quark phase $\alpha$ 
(see Appendix \ref{app_mutot_lghifc}). For HIAS\_fc, the hadronic phase has a higher baryon
chemical
potential than the quark phase, $\mu_B^H>\mu_B^Q$. Therefore an increase of the
baryon number density inside the phase coexistence region will lead to a
decrease of $\mu_B$. For $\alpha=1$, i.e.\ $\rho_B=\rho_B^Q$, the CC in
Fig.~\ref{fig:03TxMu} is reached. For even higher densities
$\mu_B$ will increase again.

\begin{figure}
\begin{center}
\includegraphics[width=\columnwidth]{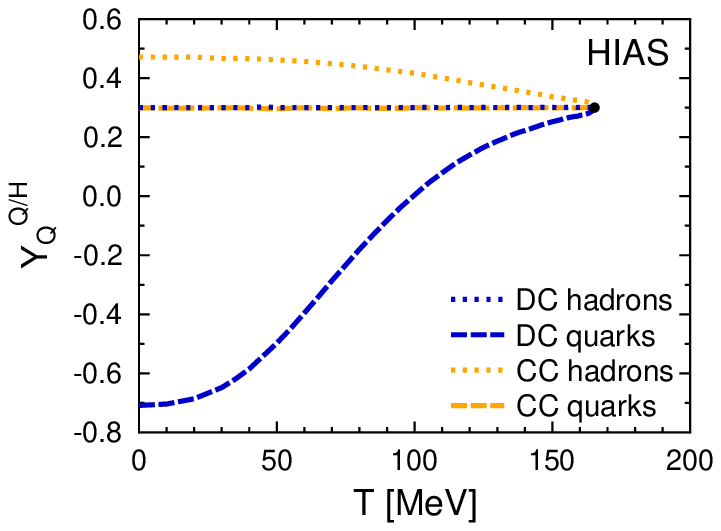}
\caption{\label{fig:03YcxT}(Color online) 
The charge fractions of the hadronic ($Y_Q^H$, dotted lines) and the quark phase
($Y_Q^Q$, dashed lines) as a function of the coexistence temperature along the
phase boundary, for case HIAS. The blue lines show states on the
deconfinement curve DC, and the orange on the confinement curve CC.}
\end{center}
\end{figure}
\begin{figure}
\begin{center}
\includegraphics[width=\columnwidth]{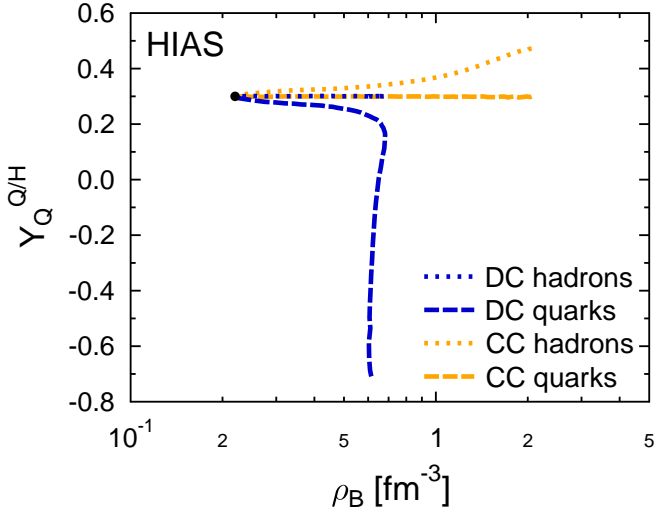}
\caption{\label{fig:03YcxRho}(Color online) As Fig.~\ref{fig:03YcxT}, but as a function of
the coexistence baryon density.}
\end{center}
\end{figure}

Figures \ref{fig:03YcxT} and \ref{fig:03YcxRho} 
show the charge fraction of each
phase (hadronic and quark) as a function of the temperature and the baryon
density along the phase boundaries.  For clarity, we again distinguish states on the deconfinement
curve by blue color and states on the confinement curve by orange.
In Figures \ref{fig:03YcxT} and \ref{fig:03YcxRho} one sees that
the charge fraction in the quark phase is always less than or equal to the
charge fraction in the hadronic phase. This shows that the charge fraction
is also an order parameter of the QHPT. Note that $Y_Q^Q\leq Y_Q^H$ 
is the opposite behavior compared to the liquid-gas PT, in which the denser 
phase is more symmetric. 

Note also that
$Y_Q$ can in principle take negative values for both the hadronic and the quark
phase, due to negatively charged hyperons, respectively down and strange quarks.
In the LGPT we only considered neutrons and protons, so that $0\leq Y_Q \leq 1$.
For HIAS we have $-1\leq Y_Q \leq 1$. Indeed we observe in 
Fig.~\ref{fig:03YcxT} that $Y_Q^Q<0$ on the DC
for coexistence temperatures below approximately 100~MeV. This means that the first
quark matter droplets which would appear at the deconfinement phase boundary of
the hadronic phase (DC) are \textit{negatively}
charged for such temperatures. Remember that on the DC only
$Y_Q^H$ is constrained to the value 0.3, and $Y_Q^Q$ is set by the equilibrium
conditions. Conversely, if the PT
is crossed coming from the high density side we have $Y_Q^Q=0.3$ on the CC. The
hadronic phase on the CC shows the opposite tendency than quark matter. For low
temperatures it approaches a rather symmetric configuration with $Y_Q$ close to
0.5. Similar features have been discussed, e.g., in Ref.~\cite{fischer11} for a
simple quark-bag model.
Fig.~\ref{fig:03YcxRho} appears quite complex and
looks different to the equivalent Fig.~\ref{fig:ynb0.3} of LGAS.
This can be explained by the
non-monotonous behavior of
the density as a function of the coexistence temperature in case HIAS, visible
in Fig.~\ref{fig:03TxRho}. 

\begin{figure}
\begin{center}
\includegraphics[width=0.9\columnwidth]{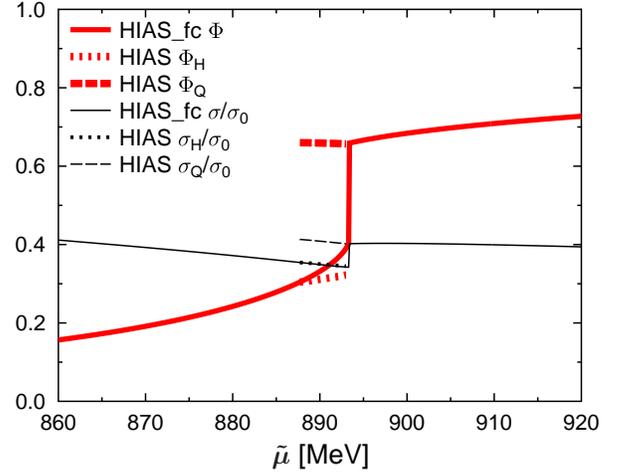}
\caption{\label{fig:03PolxMuhat}(Color online) Normalized order parameter for chiral symmetry
restoration/breaking $\sigma$ (black thin lines) and order parameter for
deconfinement/confinement $\Phi$ (red thick lines) versus the Gibbs free
energy per baryon
$\tilde \mu$ for case HIAS. Inside the phase coexistence region, the values of
the order parameters in the quark phase (dashed lines) are different from the
ones in the hadronic phase (dotted), and change during the phase transformation.
HIAS\_fc is the forced congruent variant, shown by solid lines. All order
parameters are calculated for the trajectory through the phase diagrams shown in
Fig.~\ref{fig:03TxMuhat} with $T=0.15 \tilde \mu$.}
\end{center}
\end{figure}
The two figures \ref{fig:03YcxT} and \ref{fig:03YcxRho} 
also show how the charge fraction in both phases goes to $0.3$
when approaching the critical point. At the critical point, the two phases are
identical, i.e.\ have the same charge fraction, density, scalar field $\Phi$,
chiral condensate, etc. To better understand the dynamics of such a
NCPT, in Fig.~\ref{fig:03TxMuhat} we included an
example for a trajectory through the phase diagram for which we show the order
parameters in Fig.~\ref{fig:03PolxMuhat}. The trajectory was chosen to follow
$T=0.15\tilde \mu$ so that the phase coexistence region is crossed at
temperatures around 900~MeV and $\tilde \mu \sim 130$~MeV. This trajectory could
e.g.\ be realized during the decompression of the quark-gluon plasma in a
heavy-ion collision. We show the behavior of the order parameters for HIAS and
HIAS\_fc to illustrate the differences between a non-congruent and congruent
PT. For HIAS\_fc, when the phase transition line is crossed, there
are jumps in $\sigma$ and $\Phi$, as expected for a first-order PT
(compare also with Fig.~\ref{pol}). The quark phase is characterized by having a
larger value of $\Phi$, showing that it represents the deconfined state. As can
be seen from Eqs.~(\ref{1}) and (\ref{2}), this increases the effective mass of
baryons, but decreases the effective mass of quarks. 
Note that
$\sigma$, which is decreasing with increasing $\tilde \mu$ in
both of the two phases, locally increases at the PT
(going from left to the right), because the value of $\sigma$ is higher in
the quark phase. 
This is in contrast
to the typical expectation of other models which give
partial chiral symmetry restoration, i.e., a lower value 
of $\sigma$ in the quark phase. 
This effect found in our calculations comes from the fact that in the 
Chiral SU(3) model both order parameters $\sigma$ and $\Phi$ are connected 
through the effective masses of the particles.
Also the baryonic density of the quarks (not counting the contribution from
$U$) is less than the one of the hadrons. This
leads to a decrease of the scalar field across the phase
transition. 

In Fig.~\ref{fig:03PolxMuhat} it can be seen that in the non-congruent
phase transition HIAS the behavior of the order parameters is more complex.
Within the phase coexistence region, i.e.\ within the DC and CC shown in
Fig.~\ref{fig:03TxMuhat}, the hadronic and the quark phases are in coexistence.
The two phases are spatially separated and in each of the two phases one has
different values of the order parameters. At the onset of the PT
at the confinement curve one has $\alpha=1$, i.e.\ the volume fraction of the
quark phase is still one and the volume fraction of the hadronic phase is 0.
Inside the phase coexistence region, the quark phase has a larger value of
$\Phi$, but also a slightly increased chiral condensate $\sigma$, as observed
for HIAS\_fc before. With decreasing $\tilde \mu$, not only the volume fraction
of the quark phase decreases, but also the properties of the two phases change.
One clearly sees that chiral symmetry breaking proceeds in both of the two
phases with decreasing density. On the other hand $\Phi^H$ is decreasing, and
$\Phi^Q$ is slightly increasing. When the deconfinement curve is reached,
$\alpha=0$ and only the hadronic phase is left. After this in Fig.~\ref{fig:03PolxMuhat}
one sees that $\Phi^H$ and $\sigma^H$ become equal to values obtained for
HIAS\_fc. 

\begin{figure}
\begin{center}
\includegraphics[width=\columnwidth]{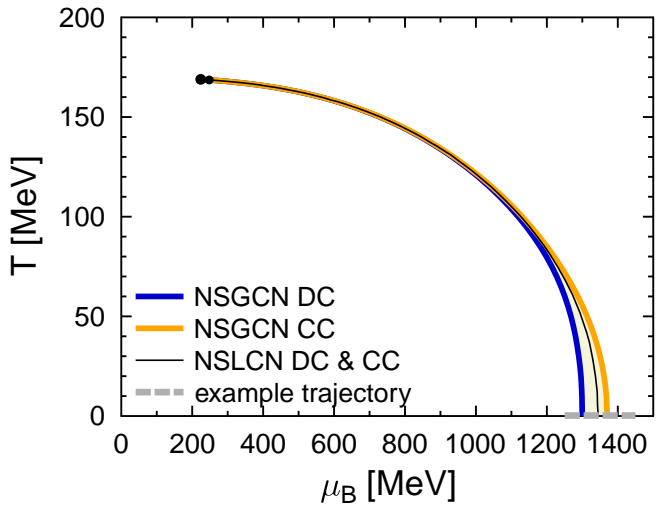}
\caption{\label{fig:NSTxMu}(Color online) Phase diagram in the temperature - baryon chemical
potential plane for neutron stars with local (NSLCN) and global charge neutrality (NSGCN). The horizontal thick gray line
at $T=0$ is a trajectory through the phase diagram, for which we show the order
parameters in Fig.~\ref{fig:NSPolxMu}. The large black dot is the
critical point of NSGCN, the small one the pseudo-critical point of
NSLCN.}
\end{center}
\end{figure}
Finally, we discuss results for the quark-hadron phase transition in neutron
stars shown in Fig.~\ref{fig:NSTxMu}. 
Let us reiterate that in this paper we consider the quark-hadron PT in
multi-component matter of neutron stars of macroscopic coexisting
phases (i) with local charge
neutrality and (ii) in the variant of global charge neutrality within the
Coulomb-less approximation. It is just the
Coulomb-less approximation
which justifies the use of Gibbs conditions instead of
Gibbs-Guggenheim
conditions, which are valid for true macroscopic Coulomb systems (e.g.,
Ref.~\cite{iosilevskiy10}).
Local and global charge neutrality
constraints lead to forced-congruent (``Maxwell'') and non-congruent (``Gibbs'')
PTs. As discussed in Sec.~\ref{sec_coul},
the choice between local or global charge neutrality can be
associated with the
unknown surface tension between the two phases.

The phase diagram of neutron stars looks similar to the one shown in
Fig.~\ref{fig:03TxMuhat}. In contrast, the dimensionality of HIAS\_fc in
Fig.~\ref{fig:03TxMu} is different to the one of NSLCN in
Fig.~\ref{fig:NSTxMu}. This is easy to understand, because 
$\mu_B^Q=\mu_B^H$ is valid for NSLCN and NSGCN, and also for HIAS, but not for
HIAS\_fc. 
The critical points of the neutron star cases are at lower
$\mu_B$ and slightly higher $T$ than in cases HIAS and HIS, i.e., the phase
transition regions extend to higher temperatures. Asymmetry and
charge neutrality do not seem to have a major effect at such high
temperatures. Therefore we relate this difference to the treatment of
strangeness (see
Table~\ref{tab:cases}). At
$T=0$ the width of the phase coexistence region of HIAS is 52~MeV, for
NSGCN it is 70.2~MeV.  With the chiral SU(3) model used
here, the hadron-quark PT and the non-congruent features
at $T=0$ seem to
be stronger for neutron star matter than for matter in heavy-ion collisions.
This is a result of the larger asymmetries obtained in beta-equilibrium
due to the large degeneracy of electrons, and also because of the
different constraints used for strangeness.

The critical point of NSGCN is at approximately $T^{\rm CP}=168.9$~MeV
and $\mu_B^{\rm CP}=224$~MeV, the
one of NSLCN at $T^{\rm CP}=168.6$~MeV and $\mu_B^{\rm CP}=247$~MeV (see
also Table~\ref{tab:cps}). Interestingly, they are only slightly different.
It shows that the treatment of electric charge neutrality plays only a minor
role for the location of critical points of the QHPT at the typical high
temperatures. The local constraint applied in HIAS\_fc which does not allow
isospin diffusion, has a slightly larger effect at high temperatures.

The phase transition line for NSLCN with enforced local charge neutrality
must lie strictly
within the phase coexistence region of NSGCN, in accordance with general rules
for NCPTs (e.g.,
Refs.~\cite{I1,I2,I3,I4,I5,iosilevskiy03}). Both boundaries can touch each other
in points of azeotropic composition. Here this would mean that the two
phases were charge neutral a priori. Actually this is not the case, but
they only touch because of the thickness of the lines in the figure.
For NSLCN, $\mu_Q$ will not behave
continuously when the phase transition line is crossed, as already discussed at
the end of Sec.~\ref{sec_eq}, because $\mu_Q^H$ is different from $\mu_Q^Q$.
This is in accordance with general properties of phase coexistence of
charge neutral, macroscopic phases in Coulomb systems (see, e.g.,
Refs.~\cite{iosilevskiy10,I5,iosi00} and
\cite{aqua12}): any phase-interface in macroscopic equilibrium Coulomb
systems is accompanied by a finite difference in the
average electrostatic potentials of both coexisting phases (Galvani potential),
see also Sec.~\ref{sec_coul} and compare also with Ref.~\cite{ebel10}. 
However, the total charge chemical potential $\mu_Q$ still can be related to the
local chemical potentials $\mu_Q^H$ and $\mu_Q^Q$, if it is seen as a function
of baryon number density. Based on two different assumptions about the
implementation of local charge neutrality, in Appendix \ref{app_mutot_ns} we
derive the following expressions:
\begin{eqnarray}
\mu_Q = \begin{cases}
         (1-\alpha) \mu_Q^H + \alpha \mu_Q^Q, & \mbox{for } \rho_Q^H \equiv
\rho_Q^Q \\
         (1-\alpha) \frac{\rho_B^H}{\rho_B}\mu_Q^H + \alpha
\frac{\rho_B^Q}{\rho_B}\mu_Q^Q, & \mbox{for } Y_Q^H\equiv Y_Q^Q
        \end{cases}
\end{eqnarray}
It would be more intuitive to assume the first of the two conditions. These
expressions can be used to determine the total chemical potential of
charged particles
inside the two-phase mixture.

\begin{figure}
\begin{center}
\includegraphics[width=0.9\columnwidth]{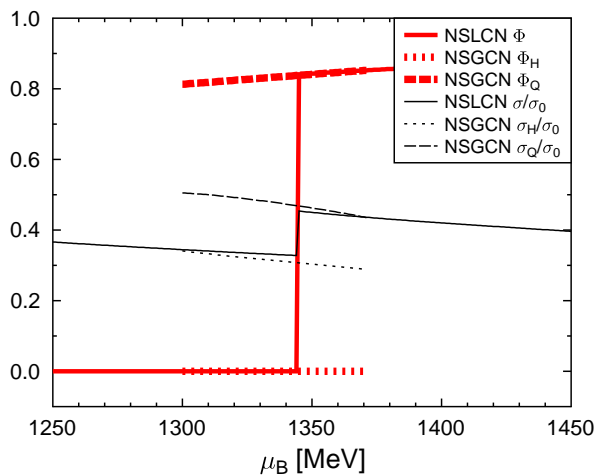}
\caption{\label{fig:NSPolxMu}(Color online) As Fig.~\ref{fig:03PolxMuhat}, but now for cases
NSLCN and NSGCN of the example trajectory at $T=0$.}
\end{center}
\end{figure}
Fig.~\ref{fig:NSTxMu} also includes an example trajectory through the phase
transition region at $T=0$. This trajectory could correspond e.g.\ to the
spatial structure inside a neutron star, or better called hybrid star. In
Fig.~\ref{fig:NSPolxMu} we show the behavior of the order parameters for cases
NSLCN and NSGCN along the trajectory. For $T=0$, we find $\Phi^H=0$,
i.e.\ the hadronic phase is completely confined. For finite temperature this is
different as can be seen in Figs.~\ref{pol} and \ref{fig:03PolxMuhat}, where one
has some sequential deconfinement already in the hadronic phase. Otherwise
similar features are observed as in Fig.~\ref{fig:03PolxMuhat}. The
obtained jumps show that we have a first-order PT. The different
values of the order parameters distinguish the two spatially separated phases.
If the PT is non-congruent, one obtains a phase coexistence
region, and the phases change their properties, including the order parameters,
during the phase transformation.  

\section{Summary and Conclusions}
\label{sec_conclusions}
In this article we investigated the liquid-gas phase transition (LGPT) described
by the FSUgold relativistic mean-field model and the hadron-quark (HQ) or
deconfinement phase transition with the Chiral SU(3) model. We
did not take into account any explicit Coulomb interactions or other finite-size
effects, but always
considered phase coexistence of macroscopic phases. Different
physical systems were investigated: heavy ion collisions
of symmetric nuclei at
low (LGS) and high collision energies (HIS) and of asymmetric nuclei with
$Z/A=0.3$ at low (LGAS) and high collision energies (HIAS). Furthermore, we also
studied the QHPT in neutron stars (NS). The main goal of our work is to
characterize the first-order phase transitions in the different systems
regarding their possible non-congruence and to identify the typical
non-congruent aspects. This characterization and classification, and our
qualitative results should not depend on the model used for the calculations,
but should be valid in a rather general way. In our study of non-congruent
phase transitions we used essential features of such type of phase transitions
obtained from terrestrial applications with high-temperature chemically reacting
plasmas \cite{I1,I2,I3,I4,I5,iosilevskiy03}.

\begin{table}
\begin{center}
\caption{\label{tab:cps}Approximate location of the critical and
pseudo-critical points of the different scenarios.}
\begin{tabular}{ccccc}
\hline
\hline
case & $T$ [MeV] & $\rho_B$ [fm$^{-3}$] & $\mu_B$ [MeV]& $p$ [MeV/fm$^{3}$]\\
\hline
LGS & 14.75 & 0.046 & 912.4 & 0.205 \\
LGAS & 13.99 & 0.049 & 927.4 & 0.241 \\
LGAS\_fc & 12.68 & 0.044 & 928.3 & 0.171 \\
HIS & 165.5 & 0.24 & 383 & 73.1 \\
HIAS & 165.3 & 0.22 & 381 & 68.6 \\
HIAS\_fc & 156.5 & 0.44 & 611 & 95.8 \\
NSLCN & 168.6 & 0.16 & 247 & 162.8 \\
NSGCN & 168.9 & 0.14 & 224 & 161.1 \\
\hline
\hline
\end{tabular}
\end{center}
\end{table}
A non-congruent, first-order phase transition occurs for a
non-unary system with more than one (globally) conserved charge where the
charge ratios change during a phase transformation. In principle, all the
investigated scenarios contain several
conserved charges, with at least baryon number and electric charge. 
However, due to isospin symmetry, for the
symmetric case (LGS and HIS), matter always stays symmetric,
even if there is phase-coexistence, which is called ``azeotropic'' behavior.
Only these two cases lead to congruent phase transitions. To further illustrate
the differences between a congruent and a non-congruent phase transition, we
also considered the cases LGAS\_fc and HIAS\_fc, where we forced the system to
have a congruent phase transition, by constraining the local charge fractions to
be equal in the two phases. From a physical point of view the
forced-congruent regime of phase transformations is not completely
artificial, but corresponds to the so-called ``frozen diffusion
approximation''. In a similar way we considered enforced local charge
neutrality in neutron stars in case NSLCN instead of global charge neutrality
NSGCN. In Table~\ref{tab:class} we summarize the classification of the phase
transitions in the different physical systems.

In the astrophysics community, congruent and non-congruent phase transitions
are usually called ``Maxwell'' and ``Gibbs'' phase transitions, referring to the
way in which phase 
coexistence is calculated. Because of the constant
charge ratios and the fact that all properties of the two phases do not change for an isothermal
phase transformation, a simple Maxwell construction can be used for the
congruent case. Conversely, for non-congruent phase transitions the more
complicated Gibbs construction is necessary to solve the set of thermal,
mechanical and multiple chemical equilibrium conditions. Due to their
prominent role in neutron star physics, NSLCN and NSGCN are considered as the
two most representative scenarios. However, as we demonstrated, they are only examples
illustrating the further classification of first-order phase transitions to be
either congruent or non-congruent.

The distinction between congruent and non-congruent phase transitions is
crucial for the dimensionality of phase diagrams. For a given temperature $T$,
a congruent phase transition occurs at a single value of the Gibbs
free energy per baryon $\tilde \mu$. Therefore in the
$T$-$\tilde \mu$-plane, one
obtains a phase transition line, as well as in the $T$-$P$-  and
$P$-$\tilde \mu$-planes, and generally in any pair of intensive thermodynamic
variables \cite{iosilevskiy10}.
 Conversely, for a non-congruent phase transition
calculated with the Gibbs construction and a given chemical composition, in any pair of intensive
thermodynamic variables one obtains two-dimensional ``banana-like''
phase-coexistence regions (see, e.g., Fig.~1 in Ref.~\cite{iosilevskiy10}).
One of the conclusions from the present study is that all asymmetric systems
are non-congruent, if no additional local constraints are enforced. The
corresponding phase diagrams should always be regions, and not lines.

{\setlength{\tabcolsep}{4pt}
\begin{table}[t]
\begin{center}
\caption{\label{tab:class}Classification of the equilibrium conditions and 
type of first-order phase transition in the different scenarios.}
\begin{tabular}{ccc}
\hline
\hline
case & equilibrium conditions & type of PT \\
\hline
 LGS  & Gibbs/Maxwell (equiv.) & congruent (azeotrope) \\
LGAS  & Gibbs & non-congruent\\
LGAS\_fc & Maxwell & forced congruent \\
 HIS  &  Gibbs/Maxwell (equiv.) & congruent (azeotrope) \\
HIAS  & Gibbs & non-congruent\\
HIAS\_fc & Maxwell & forced congruent \\
NSLCN & Maxwell & forced congruent \\
NSGCN & Gibbs & non-congruent \\
\hline
\hline
\end{tabular}
\end{center}
\end{table}
}

Inside the phase coexistence region, the order parameters (e.g.\ density and
charge fraction for LG, density, charge fraction, chiral condensate, and $\Phi$-field
representing the Polyakov-loop for QHPT) have different values in the two
phases, which is the actual definition of a first-order phase transition. We
demonstrated the behavior of some of these order parameters on the 
binodal boundaries and within two-phase regions, and how they
become equal at the critical point. Furthermore,
for the QHPT this was also illustrated for the chiral condensate and the
$\Phi$-field for trajectories through the phase diagram in a high energy
heavy-ion collision and in a neutron star. For a congruent phase transition, the
different values appear as a jump when crossing the phase transition line and
moving from one phase to the other. For non-congruent phase transitions, at the
phase coexistence boundary the second phase appears, and its volume fraction
increases until the first phase has disappeared completely when the phase
coexistence region has been crossed. During such a non-congruent phase
transformation, both phases continuously change their thermodynamic properties and order
parameters. 

Local charge neutrality for macroscopic phase coexistence in neutron stars
under Maxwell conditions has been applied
in a number of published works. It is well known, that the local charge
chemical potentials behave discontinuously in such a transition, while the
generalized non-local electro-chemical potentials are continuous in accordance with
generalized Gibbs-Guggenheim conditions (e.g.\ Ref.~\cite{iosilevskiy10}). 
The discontinuity is associated with the so-called Galvani potential of the
phase interface in Coulomb systems (e.g., Refs.~\cite{I5} and \cite{iosi_eltp2004}).
In the present article, for the Coulombless approximation we derived
expressions for the total charge chemical potential $\mu^Q$ inside
the phase coexistence region as a function of the total baryon number
density, which behaves continuously. The
new expressions show explicitly that the total chemical potentials on the
binodal surface correspond to the ones of the dominating phase if its volume
fraction is unity. We also derived the total chemical potentials for
the other phase transitions where local constraints were applied. 

Another fundamental aspect of non-congruent phase
transitions is the location and properties of
the critical point. For congruent phase transitions, the critical point is
located at the end of the phase transition line. Conversely,
the phase
coexistence region of a non-congruent phase transition has 
several (topological) endpoints: namely for temperature, pressure and chemical
potential,
which are all different (e.g., Ref.~\cite{iosilevskiy10}). The critical
point, 
 as the point of thermodynamic singularity, and where the
two phases are identical, does not
coincide with these endpoints. For the LGPT
we found that the temperature endpoint is located on the saturation curve, i.e.\
the critical point is located at higher densities than the temperature endpoint,
while the pressure endpoint is located on the boiling curve. This is
the same arrangement as for non-congruent phase transitions in terrestrial
chemically reacting plasma (e.g., Fig.~1 in Ref.~\cite{iosilevskiy10}).
The properties of all (pseudo-) critical points found in the present study
are listed in Table~\ref{tab:cps}.

For the QHPT we could not resolve the endpoints and the critical points because
we reached the limit of our numerical accuracy. But at the same time,
 this is related to another
important finding from our study: at low temperatures, the asymmetry has a
significant impact on the phase diagrams of the QHPT. Conversely, for high
temperatures, close to the critical point, this impact is vanishingly small
because the EOS is dominated by thermal contributions. As a
consequence, the non-congruent features of HIAS are almost unnoticable close to
the critical point. This is in contrast to LGAS of
nuclear matter considered here and LGPT in terrestrial chemically
reacting plasma \cite{I1,I2,I3,I4,I5,iosilevskiy03} for which the
non-congruence is significant for all conditions. 

Another important difference between LGPT and QHPT comes from the phase
diagrams in the temperature-pressure plane. For ordinary
Van-der-Waals-like phase transitions of liquid-gas
type, and also for the numerous variants of LGPT in nucleonic matter
studied in other papers, the slope of the $P$-$T$-phase
transition line is positive, which we also confirm in our study.
Conversely, for the QHPT we found that the slope is negative,
which leads to the conclusion that this phase transition is not of liquid-gas
type. The Clapeyron-equation shows that this peculiar feature is a result of the
generally higher entropy per baryon of the quark phase than of the hadronic
phase during phase coexistence. We also found that the denser phase of the QHPT
(the quark phase) is more asymmetric than the more dilute phase (the hadronic phase), 
which is also opposite to the behavior in the LGPT.

In the future, we could extend our work to study phase diagrams of
proto-neutron stars and supernovae. Those objects are special, not only for
having higher temperature in comparison to neutron stars, but also for having a
higher fixed lepton content due to the presence of trapped neutrinos. Such
features transform an already binary phase
transition into a ternary one due to the
introduction of the lepton number conservation, with potentially interesting
results. It was already shown that a phase transition to deconfined matter prior
to the supernova explosion can have observable effects \cite{sagert09}.
Another aspect which we did not discuss in detail are the consequences of
different constraints regarding strangeness conservation and possible
strangeness distillation. 

It would be interesting to explore further the 
experimental consequences of non-congruence in heavy-ion collisions.
In our investigations we have found that the non-congruence of the QHPT increases with
decreasing temperature. Therefore, the possible non-congruent features could
become particularly relevant for the heavy-ion experiments at the future FAIR
facility at Darmstadt or NICA at Dubna, which both aim to probe asymmetric matter at 
high densities and low temperatures.

\subsection*{Acknowledgments}
The authors thank Michael Strickland for proofreading the manuscript. 
I.I.\ acknowledges also V.Fortov, D.Blaschke and J.Randrup for support and fruitful discussions.
M.H.\ acknowledges support from the High Performance and High Productivity
Computing (HP2C) project and the Swiss National Science Foundation (SNF) under
Grant No. 200020-132816/1. This work has been supported additionally by
CompStar, a research
networking program of the European Science Foundation. M.H.\ is also grateful
for participating in the EuroGENESIS collaborative research program of the ESF
and the ENSAR/THEXO project. I.I.\ acknowledges support from the Scientific
Program ``Physics of extreme states of matter'' of the Russian Academy of
Science and Education Center ``Physics of high energy density matter'' of the
Moscow Institute of Physics and Technology. The work of I.I.\ was also partially
supported by the Extreme Matter Institute - EMMI (Germany).

\appendix
\section{Critical points in a ``two-EOS'' description}
\label{app_cp2eos}
It is well-known (e.g., Ref.~\cite{landau69}) that in a ``two-EOS'' description
of PTs, where two
different EOS models are used for the two phases in coexistence, there
cannot be a termination point or critical-point of a phase
transition line. This is due to the standard argument of the impossibility
of a continuous and smooth
transformation from one phase into another in a two-EOS approach (e.g.,
commonly used for crystal-fluid coexistence). This is in
contrast to the claims in the recent studies of, e.g.,
Refs.~\cite{2009PhRvD..80k4508S,2010PhRvD..82a4023S,Shao:2011fk,shao12}. The
previous statement can be supported
with the following geometrical argumentation. For symmetric
matter, the two phases give two different planes in the parameter-space of
pressure $P$, temperature $T$ and baryon chemical potential $\mu_B$. Where the
two planes intersect, one has phase coexistence. This
intersection of two two-dimensional planes in a three-dimensional parameter
space either has to be a closed or an infinite curve. A termination point of the
intersection curve is impossible, as long as the planes do not show any
discontinuities, which would be rather unphysical. This argumentation can also
be generalized to asymmetric matter, where the charge chemical potential $\mu_Q$
appears as another dimension.

\section{Azeotropic behavior for symmetric matter in HIS and LGS}
\label{app_azeo}
The azeotropic behavior of HIS and LGS can be explained as a result of isospin
symmetry in the following way. First of all, it is important to note that from
$Y_Q=0.5$ it follows that the total third component of the isospin is zero,
$I_3=0$, if there is no net strangeness, $S=0$. Furthermore, the considered
particles (neutrons and protons for LGS, u-, d-, s-quarks and the baryon octet
for HIS) can be grouped in pairs of isospin partners, where the isospin
partners
have identical baryon number, strangeness and mass, but opposite third component
of the isospin, and some remaining particles with zero
third component of the
isospin. 
If, in addition, the interactions are built in an isospin symmetric way,
namely that they are identical for the isospin partners at $I_3=0$,
it follows that $I_3=0$ if and only if the isospin
chemical potential is also zero,
$\mu_{I_3}=\frac{\partial F}{\partial I_3}=0$. For the constraints used in
LGS and HIS all these conditions are fulfilled by the two applied EOS models
FSUgold and Chiral SU(3).
On the other hand, it can be shown that the charge chemical
potential is equal to the isospin chemical potential, $\mu_{I_3}=\mu_Q$. Thus, we
obtain that $\mu_Q=0$, if $Y_Q=0.5$ and $S=0$, but independent of
temperature and density. For two-phase
coexistence in HIS and LGS we use the constraints
$S^I=S^{II}=0$ and $Y_Q=0.5$. Therefore we get $\mu_Q^I=\mu_Q^{II}=0$ and
$Y_Q^I=Y_Q^{II}=0.5$, which shows that the two phases
remain symmetric and that the PT is therefore azeotropic.

\section{Total chemical potentials inside the phase-coexistence regions}
\label{app_mutot}
In all HI and LG cases local constraints are applied for two-phase
coexistence. Therefore, it is necessary to refine the definition of the total
chemical potentials inside the two-phase mixture, and to relate them to the
local chemical potentials already defined in Table \ref{tab:consquant}.

\subsection{HIAS\_fc and LGAS\_fc}
\label{app_mutot_lghifc}
Let us start with cases LGAS\_fc and HIAS\_fc. The constraints listed in Table
\ref{tab:cases} are equivalent to considering $B={\rm const.}$,
$Q={\rm const.}=Y_QB$, $Y_Q^I=Y_Q^{II}$, and $S^I=S^{II}=0$. Note that this set
of constraints also gives $Y_Q=Y_Q^I=Y_Q^{II}$. The constraint for zero local
strangeness, $S^I=S^{II}=0$, is formulated with local extensive variables.
However, it would be equivalent to consider
$Y_S^I=Y_S^{II}$ and $S=0$, or
$\rho_S^I=\rho_S^{II}$ and $S=0$, where we define $Y_S^I=S^{I}/B^{I}$,
$\rho_S^I=S^I/V^I$, and analogous expressions for phase $II$. In conclusion, we
have the simple conservation of the total charges, which are fixed to some
values, plus two additional local constraints $Y_Q^I=Y_Q^{II}$ and
$Y_S^I=Y_S^{II}$, or $Y_Q^I=Y_Q^{II}$ and $\rho_S^I=\rho_S^{II}$. Let us first
use the first of the two formulations of the two local constraints.

Now we can define the total baryon chemical potential inside the two-phase
mixture as the following derivative
\begin{eqnarray}
 \mu_B &=& \left.\frac{\partial
F}{\partial B}\right|_{T,V,S,Q,Y_Q^I=Y_Q^{II},Y_S^I=Y_S^{II}} . \label{eqapp1}
\end{eqnarray}
This is nothing but the definition of $\mu_B$ outside the two-phase mixture
given in Table \ref{tab:qn_single}, but with the additional local constraints
taken into account. This is what we mean by the total chemical potentials
inside the two-phase mixtures. 
 For the definition of $\mu_Q$, ``$Q$'' and
``$B$'' have to be exchanged in Eq.~(\ref{eqapp1}), for $\mu_S$, ``$S$'' and
``$B$''. For $\tilde \mu$, only ``$Q$'' has to be replaced by ``$Y_Q$'' in the
list of constant variables. Note that this definition of $\tilde
\mu$ gives the same relation 
\begin{equation}
\tilde \mu = Y_Q \mu_Q + \mu_B  \label{tildemutot}
\end{equation}
obtained previously for single phases. Let us introduce $\bf X$ as an abbrevation for
the set of variables kept constant and the local constraints
\begin{eqnarray}
 \mu_B &=& \left.\frac{\partial
F}{\partial B}\right|_{\bf X} . 
\end{eqnarray}
For the definition of the total chemical potentials
of other cases than HIAS\_fc and LGAS\_fc, one only has to exchange the local
constraints in $\bf X$. 

Now we want to evaluate the above expression for
$\mu_B$ in HIAS\_fc and LGAS\_fc. Inside the two-phase mixture, the total free
energy $F$ is given as the
sum of the two phases        
\begin{equation}
 F = F^I(T,V^I,B^I,S^I,Q^I)  + F^{II}(T,V^{II},B^{II},S^{II},Q^{II}) \; ,
\end{equation}
where we used thermal equilibrium, Eq.~(\ref{condeqg2}). With the chain rule
and the definitions of the local chemical potentials of
Table~\ref{tab:consquant} we obtain
\begin{eqnarray}
 \mu_B &=& 
-\left.\frac{\partial V^I}{\partial B}\right|_{\bf X}P^I
-\left.\frac{\partial V^{II}}{\partial B}\right|_{\bf X}P^{II}  \nonumber \\ 
&& +\left.\frac{\partial B^I}{\partial B}\right|_{\bf X}\mu_B^I
+\left.\frac{\partial B^{II}}{\partial B}\right|_{\bf X}\mu_B^{II} \nonumber \\
&& +\left.\frac{\partial S^I}{\partial B}\right|_{\bf X}\mu_S^I
+\left.\frac{\partial S^{II}}{\partial B}\right|_{\bf X}\mu_S^{II} \nonumber \\
&& +\left.\frac{\partial Q^I}{\partial B}\right|_{\bf X}\mu_Q^I
+\left.\frac{\partial Q^{II}}{\partial B}\right|_{\bf X}\mu_Q^{II} \; .
\end{eqnarray}
Because of pressure equilibrium,
Eq.~(\ref{condeqg1}), the first two terms sum up to zero. Next we use
$S^I=Y_S^I B^I$ for the expression $\left.\frac{\partial S^I}{\partial
B}\right|_{\bf X}$
\begin{eqnarray}
 \left.\frac{\partial S^I}{\partial B}\right|_{\bf X} =
 Y_S^I \left.\frac{\partial B^I}{\partial B}\right|_{\bf X}
+ B^I \left.\frac{\partial Y_S^I}{\partial B}\right|_{\bf X} = B^I
\left.\frac{\partial Y_S^I}{\partial B}\right|_{\bf X} , \; \; \; \;  \;
\end{eqnarray}
where the second equality comes from $Y_S^I=0$. If we have $Y_S^I=Y_S^{II}$, we
also have $Y_S^I=Y_S^{II}=Y_S$, and thus
\begin{eqnarray}
B^I\left.\frac{\partial Y_S^I}{\partial B}\right|_{\bf X} = B^I
\left.\frac{\partial
Y_S}{\partial B}\right|_{\bf X} = -\frac{B^I}{B}Y_S = 0 \; .
\end{eqnarray}
In conclusion, we obtain $\left.\frac{\partial S^{I}}{\partial
B}\right|_{\bf X}=0$, and in the same way also $\left.\frac{\partial
S^{II}}{\partial
B}\right|_{\bf X}=0$. Thus we are left with
\begin{eqnarray}
 \mu_B =
&& \left.\frac{\partial B^I}{\partial B}\right|_{\bf X}\mu_B^I
+\left.\frac{\partial B^{II}}{\partial B}\right|_{\bf X}\mu_B^{II} \nonumber \\
&+& \left.\frac{\partial Q^I}{\partial B}\right|_{\bf X}\mu_Q^I
+\left.\frac{\partial Q^{II}}{\partial B}\right|_{\bf X}\mu_Q^{II} 
\; .
\end{eqnarray}
To make use of the local constraint $Y_Q^I=Y_Q^{II}$ in $\bf X$, we replace
$Q^I$ by $Y_Q^IB^I$, and $Q^{II}$ by $Y_Q^{II}B^{II}$, and use
$Y_Q=Y_Q^I=Y_Q^{II}$. This gives
\begin{eqnarray}
 \mu_B =
&& 
\left.\frac{\partial B^I}{\partial B}\right|_{\bf X}
(\mu_B^I+Y_Q^I\mu_Q^I) \nonumber \\
&+&\left.\frac{\partial B^{II}}{\partial B}\right|_{\bf X}
(\mu_B^{II}+Y_Q^{II}\mu_Q^{II}) \nonumber \\
&-& \frac{B^I}{B}Y_Q^I\mu_Q^I -\frac{B^{II}}{B}Y_Q^{II}\mu_Q^{II}\; .
\end{eqnarray}
Now we can use the inter-phase equilibrium conditions (\ref{eq_eq1}) and
(\ref{eq_eq}), and the definitions of $\tilde\mu^I$ and $\tilde\mu^{II}$,
(Eq.~(\ref{def:mut})), to obtain 
\begin{eqnarray}
 \mu_B =
&& \tilde \mu^I - \frac{B^I}{B}Y_Q^I\mu_Q^I -\frac{B^{II}}{B}Y_Q^{II}\mu_Q^{II}
\nonumber \; , \\
= && \frac{B^I}{B}\tilde \mu^I +\frac{B^{II}}{B}\tilde \mu^{II}
-\frac{B^I}{B}Y_Q^I\mu_Q^I -\frac{B^{II}}{B}Y_Q^{II}\mu_Q^{II} \nonumber \; , \\
= && \frac{B^I}{B} \mu_B^I +\frac{B^{II}}{B}\mu_B^{II} \; .
\end{eqnarray}

This can be written as
\begin{eqnarray}
 \mu_B &=& (1-\alpha) \frac{\rho_B^I}{\rho_B}\mu_B^I + \alpha
\frac{\rho_B^{II}}{\rho_B}\mu_B^{II} \; , \label{eqmub}
\end{eqnarray}
where $\alpha$ is the volume fraction of phase $II$ in the two-phase
mixture
\begin{equation}
 \alpha = \frac{\rho_B-\rho_B^I}{\rho_B^{II}-\rho_B^I} \; .
\end{equation}
For $\rho_B=\rho_B^I$ one gets $\mu_B=\mu_B^{I}$, and for $\rho_B=\rho_B^{II}$
$\mu_B=\mu_B^{II}$, i.e.\ the correct limits are obtained. Furthermore, if
$\mu_B^I=\mu_B^{II}$ one finds that $\mu_B=\mu_B^I=\mu_B^{II}$, independent of
$\rho_B$. With a similar derivation one obtains the following expression for the
total charge chemical potential
\begin{eqnarray}
 \mu_Q 
&=& (1-\alpha) \frac{\rho_B^I}{\rho_B}\mu_Q^I + \alpha
\frac{\rho_B^{II}}{\rho_B}\mu_Q^{II} \; . \label{eqmuq}
\end{eqnarray}
It turns out that one obtains the same expressions (\ref{eqmub}) and
(\ref{eqmuq}), if one uses the strangeness constraint in the form
$\rho_S^I=\rho_S^{II}$ instead of $Y_S^I=Y_S^{II}$. Furthermore, note that
Eqs.~(\ref{eqmub}), (\ref{eqmuq}), and (\ref{tildemutot}) also allow one to express
$\tilde \mu$ in terms of the local chemical potentials.

Interestingly, for the total strange chemical potential the form of the
strangeness constraint makes a difference
\begin{eqnarray}
\mu_S = \begin{cases}
         (1-\alpha) \mu_S^I + \alpha \mu_S^{II}, & \mbox{for }
\rho_S^I=\rho_S^{II} \\
         (1-\alpha) \frac{\rho_B^I}{\rho_B}\mu_S^I + \alpha
\frac{\rho_B^{II}}{\rho_B}\mu_S^{II}, & \mbox{for } Y_S^I= Y_S^{II}\; .
        \end{cases} \label{eqmus}
\end{eqnarray}
Note that the two forms give identical results if $\alpha
=0$ or if $\alpha=1$, i.e., at the phase boundaries.

Eqs.~(\ref{eqmub}), (\ref{eqmuq}), and (\ref{eqmus}) can be used to determine
the chemical potential of particles for the two-phase mixture as a whole.
For example, for protons and neutrons one obtains using Eq.~(\ref{mus})
\begin{eqnarray}
 \mu_p &=& (1-\alpha) \frac{\rho_B^I}{\rho_B}\mu_p^I + \alpha
\frac{\rho_B^{II}}{\rho_B}\mu_p^{II} \; , \\
 \mu_n &=& (1-\alpha) \frac{\rho_B^I}{\rho_B}\mu_n^I + \alpha
\frac{\rho_B^{II}}{\rho_B}\mu_n^{II} \; .
\end{eqnarray}
For $\Lambda$'s the constraint $Y_S^I=Y_S^{II}$ would, e.g., lead to
\begin{eqnarray}
 \mu_\Lambda &=& (1-\alpha) \frac{\rho_B^I}{\rho_B}\mu_\Lambda^I + \alpha
\frac{\rho_B^{II}}{\rho_B}\mu_\Lambda^{II}  \; .
\end{eqnarray}

\subsection{LGS, LGAS, HIS, and HIAS}
\label{app_mutot_lghi}
In cases LGS, LGAS, HIS, and HIAS the only local constraint is
$S^I=S^{II}=0$. Here one finds
\begin{eqnarray}
 \mu_B &=& \mu_B^I = \mu_B^{II} \; , \\
 \mu_Q &=& \mu_Q^I = \mu_Q^{II} \; , \\
\mu_S &=& \begin{cases}
         (1-\alpha) \mu_S^I + \alpha \mu_S^{II}, & \mbox{for }
\rho_S^I=\rho_S^{II} \\
         (1-\alpha) \frac{\rho_B^I}{\rho_B}\mu_S^I + \alpha
\frac{\rho_B^{II}}{\rho_B}\mu_S^{II}, & \mbox{for } Y_S^I= Y_S^{II} \; .
        \end{cases} \nonumber \\
&& 
\end{eqnarray}

\subsection{NSLCN and NSGCN}
\label{app_mutot_ns}
For case NSGCN, which uses the Coulomb-less approximation and leads to
a non-congruent PT, all three local chemical potentials have equal values
in the two phases and one obtains
\begin{eqnarray}
 \mu_B &=& \mu_B^I = \mu_B^{II} \; , \\
 \mu_Q &=& \mu_Q^I = \mu_Q^{II} \; , \\
 \mu_S &=& \mu_S^I = \mu_S^{II} (=0) \; .
\end{eqnarray}

In Table \ref{tab:cases} local charge neutrality constraint in NSLCN was
formulated as $Y_Q^I=Y_Q^{II}=0$, i.e.\ in terms of the local electric
charge fractions. Here one has a similar situation as for
the local strangeness constraint in Appendix~\ref{app_mutot_lghifc}: it would be
equivalent to consider $\rho_Q^I=\rho_Q^{II}=0$ instead of $Y_Q^I=Y_Q^{II}=0$.
After a
similar derivation as in Appendix~\ref{app_mutot_lghifc} one obtains 
\begin{eqnarray}
 \mu_B &=& \mu_B^I = \mu_B^{II} \; , \\
 \mu_S &=& \mu_S^I = \mu_S^{II} (=0) \; , \\
\mu_Q &=& \begin{cases}
         (1-\alpha) \mu_Q^I + \alpha \mu_Q^{II}, & \mbox{for } \rho_Q^I =
\rho_Q^{II} \\
         (1-\alpha) \frac{\rho_B^I}{\rho_B}\mu_Q^I + \alpha
\frac{\rho_B^{II}}{\rho_B}\mu_Q^{II}, & \mbox{for } Y_Q^I = Y_Q^{II}\; .
        \end{cases} \nonumber \\
&&
\end{eqnarray}
For the determination of $\mu_Q$ it makes a difference whether one assumes equal
charge fractions or equal charge densities in the two phases. 
It would be more intuitive to assume the first of the two conditions,
namely that additional charge is distributed uniformly, because here we have in
mind that the Coulomb forces are the reason for local charge neutrality. 

\section{Gibbs free energy}
\label{app_gibbsfe}
Let us first assume we are outside of the two-phase coexistence region,
i.e., there is only one phase. The
Gibbs free energy $G$ can
be obtained via a Legendre-transformation from the (Helmholtz) free energy $F$
\begin{equation}
 G=F+PV \; .
\end{equation}
If $G=G(T,P,\{N_i\})$ is seen as a function of the particle numbers $N_i$, which
are fixed by Eq.~(\ref{mus}), one obtains
\begin{eqnarray}
 G&=&\sum_i N_i \mu_i \\
 &=& B \mu_B + Q \mu_Q + S \mu_S \; . \label{defg}
\end{eqnarray}
This shows that $G$ can also be seen as a function of $B$, $Q$,
and $S$, i.e., $G=(T,P,B,Q,S)$. Furthermore, because we consider either
$S=0$ or $\mu_S=0$, we obtain from Eqs.~(\ref{defg}), (\ref{mutilde1}) and
(\ref{mutilde2})
\begin{eqnarray}
 G &=& B \mu_B + Q \mu_Q \\
  &=& B \tilde \mu \; ,
\end{eqnarray}
which shows that $\tilde \mu$ is indeed the Gibbs free energy per
baryon, respectively the two definitions are equivalent.

Inside the two-phase mixture we have
\begin{eqnarray}
G &=& G(T,P,\{N_i^I\},\{N_i^{II}\}) \nonumber \\
&=& G^I(T,P,\{N_i^{I}\})+G^{II}(T,P,\{N_i^{II}\}) \nonumber \\
&=& \sum_i N_i^I \mu_i^I + \sum_i
N_i^{II} \mu_i^{II} \\
&=& B^I \mu_B^I + B^{II} \mu_B^{II}+ Q^I \mu_Q^I + Q^{II} \mu_Q^{II}
\nonumber \\
&& + S^I \mu_S^I  + S^{II} \mu_S^{II} \; ,
\end{eqnarray}
where we used Eqs.~(\ref{mui_coex}). 
Interestingly, the total chemical
potentials given in Appendix~\ref{app_mutot} also fulfill
the following equalities
\begin{eqnarray}
G &=& B \mu_B + Q \mu_Q + S \mu_S \\
&=& G(T,P,B,Q,S) \; .
\end{eqnarray}
Note that 
\begin{eqnarray}
G^I(T,P,\{N_i^{I}\}) = B^I \mu_B^I + S^{I} \mu_S^{I}+ Q^I \mu_Q^I \; ,
\end{eqnarray}
and because we always consider either $S^I=S^{II}=0$ or
$\mu_S^I=\mu_S^{II}=0$, we obtain from Eq.~(\ref{def:mut})
\begin{eqnarray}
 G^I &=& B^I \mu_B^I + Q^I \mu_Q^I \\
  &=& B^I \tilde \mu^I \; ,
\end{eqnarray}
so the local $\tilde \mu^I$ is also equal to the local Gibbs free energy per
baryon.

We remark that the 
Gibbs free energy per baryon is important because it
is just this quantity which must be equal in the case of
forced-congruent phase coexistence (under Maxwell conditions), in analogy to
the specific Gibbs free energy used for terrestrial chemically
reacting plasma.

\bibliographystyle{apsrev}
\bibliography{references}
\end{document}